\documentclass[11p,a4paper]{article}
\pdfoutput=1
\usepackage{jheppub}
\usepackage[utf8]{inputenc}
\usepackage{latexsym,amsfonts,amsmath,amscd,amssymb,epsf}
\usepackage{hyperref}
\usepackage{soul,xcolor}
\usepackage{natbib}
\usepackage{enumitem}
\usepackage{graphicx}

\newcommand{\beq}{\begin{equation}}
\newcommand{\eeq}{\end{equation}}
\newcommand{\cK}{{\cal K}}
\newcommand{\cP}{{\cal P}}

\DeclareMathAlphabet{\pazocal}{OMS}{zplm}{m}{n}

\usepackage{bm}
\let\vec=\bm

\title{Medium-induced radiation with vacuum propagation in the pre-hydrodynamics phase}

\author[a]{Carlota Andres,}
\author[b,c]{Liliana Apolin\'ario,}
\author[d]{Fabio Dominguez,}
\author[d]{Marcos Gonzalez Martinez}
\author[d]{and Carlos A. Salgado}
\emailAdd{carlota.andres-casas@polytechnique.edu}
\emailAdd{liliana@lip.pt}
\emailAdd{fabio.dominguez@usc.es}
\emailAdd{marcosg.martinez@usc.es}
\emailAdd{carlos.salgado@usc.es}

\affiliation[a]{CPHT, CNRS, \'Ecole polytechnique,
IP Paris, F-91128 Palaiseau, France}
\affiliation[b]{LIP, Av. Prof. Gama Pinto, 2, P-1649-003 Lisboa, Portugal}
\affiliation[c]{Instituto Superior T\'{e}cnico (IST), Universidade de Lisboa, Avenida Rovisco Pais 1, 1049-001 Lisboa, Portugal}
\affiliation[d]{Instituto Galego de F\'isica de Altas Enerx\'ias (IGFAE), Universidade de Santiago de Compostela, Santiago de Compostela 15782, Spain}
\begin{document}

\abstract{The recent discovery of the potential of jet quenching observables to constrain the initial stages after a heavy-ion collision makes imperative to have a better understanding of the process of medium-induced radiation before the formation of the quark-gluon plasma (QGP) and its impact on observables at high-$p_T$. In this work, we generalize the BDMPS-Z framework for medium-induced radiation to account for additional emissions occurring before the creation of the QGP. For simplicity, we assume that during the pre-hydrodynamics phase the hard parton propagates as in vacuum. This set-up, allows us to isolate the contribution from the additional initial radiation by comparing with the usual scenarios in which the emitter is created inside the medium but with different starting points. Using both a numerical implementation of the fully resummed emission spectrum and the usual analytical approximations,  we find that replacing an initial slab of the medium by vacuum yields to a significant reduction of the  emission spectrum for low radiated gluon energies, while the high-energy tails remain largely unmodified.  Finally, we assess the effect of replacing the initial medium by vacuum propagation on the single-inclusive particle suppression $R_{AA}$ and high-$p_T$ azimuthal asymmetry $v_2$. Our findings indicate that considering vacuum propagation prior to hydrodynamization leads to an increase in the $v_2$, thus corroborating the importance of the treatment of jet quenching in the initial stages for the correct description of both observables.}

\maketitle

\section{Introduction}
\label{sec:intro}

The modification of jet structure in heavy-ion collisions with respect to p-p collisions, usually known as jet quenching, is one of the most useful and versatile tools to study the quark-gluon plasma (QGP) created in these collision systems -- see for instance \cite{Apolinario:2022vzg,Cunqueiro:2021wls,Connors:2017ptx,Qin:2015srf,Blaizot:2015lma,Mehtar-Tani:2013pia} and references therein. One of the main advantages of using jet quenching phenomena to probe the dynamics of the QGP is that the measured jets come from hard probes originated in the initial hard scattering which have witnessed the full evolution of the system, thus carrying information about all its different stages \cite{Apolinario:2017sob,Andrews:2018jcm,Apolinario:2020uvt}. Nevertheless, theoretical calculations often ignore the initial part of that evolution in favor of only studying the interaction between the hard probe and the plasma phase of the system, where a hydrodynamic description applies for the bulk degrees of freedom. This choice is usually justified on the grounds that the formation of the QGP is very fast and therefore the error introduced by ignoring this initial phase is not significant when calculating jet quenching observables. Recent analyses \cite{Andres:2019eus,Zigic:2019sth} have shown that this logic is misguided by pointing to the fact that in order to be able to describe the behavior of multiple jet observables simultaneously it is necessary to have a better description of the emission process in the initial stages after the collision. Furthermore, recent developments in this direction in the Color Glass Condensate context have shown that the early Glasma phase may have a significant impact on jet quenching physics \cite{Ipp:2020mjc,Ipp:2020nfu,Carrington:2021dvw,Carrington:2022bnv,Hauksson:2021okc}.

Most analytical calculations of in-medium radiation, including recent efforts \cite{Feal:2018sml,Mehtar-Tani:2019tvy,Barata:2021wuf,Andres:2020vxs,Schlichting:2021idr,Sadofyev:2021ohn,Barata:2022krd,Andres:2022ndd,Isaksen:2022pkj}, assume that the initial hard particle is created \emph{inside} the medium, hence allowing it to immediately interact with the QGP. In practice, this implies that the particle is set to be produced after the medium forms, thus ignoring the initial part of its propagation. Notice that this approximation not only ignores the effect of the strong fields present in the pre-hydrodynamics phase of the system's evolution, but also the radiation emitted before the formation of the QGP and its interference with radiation occurring inside the QGP. It is then important to highlight that the presence of this additional radiation implies that the energy loss of high-energy partons in the QGP cannot be expressed in terms of the (weighted) path length. Partons going through the same trajectories inside the QGP can lose different amounts of energy depending on the exact point where they were created.

In order to properly account for this initial propagation, one must then decouple the production time of the hard parton from the time when the medium is formed.\footnote{The case where the initial parton is allowed to radiate before entering the QGP was considered in \cite{Wiedemann:1999fq} but only for an on-shell parton created at infinity. Here we consider the creation of the parton at the collision point before the QGP is created.} In that spirit, we will then denote the production time of the initial parton as $\tau_{\rm p} \sim 1/E \ll 1$\,fm, $E$ being its initial energy, and the gluon can be emitted any time after $\tau_{\rm p}$. The time where the hydrodynamic description of the medium starts will be denoted as $\tau_{\rm m} \sim 1$\,fm and we will assume that there are no interactions before $\tau_{\rm m}$, with the initial parton propagating as in vacuum, since proposing a detailed model for the interactions in the pre-hydrodynamics phase is beyond the scope of this paper. This particular configuration allows us to isolate the contribution from this initial radiation and compare with the relevant scenarios where the initial parton is produced inside the medium but with different starting points. This is further motivated by recent findings \cite{Andres:2019eus,Zigic:2019sth} indicating that present energy loss implementations need a strong suppression of jet quenching in the pre-hydrodynamic phase to simultaneously describe the $R_{\rm AA}$ and high-$p_T$ $v_2$.

Several phenomenological studies \cite{Renk:2010qx,Armesto:2009zi,Andres:2016iys,Andres:2017awo,Andres:2019eus} have considered the issue of varying the starting time of the energy loss, thus showing some sensitivity to quenching in the initial stages. Here we aim towards determining the size of the effect arising from the initial radiation before having medium interactions by directly comparing the radiation spectrum obtained by varying the initial point of the interactions with the QGP with and without the contribution from early radiation. Specifically, we will consider the three following scenarios illustrated in figure~\ref{fig:picture_cases}:

\begin{figure}
\centering
\includegraphics[width=0.50\textwidth]{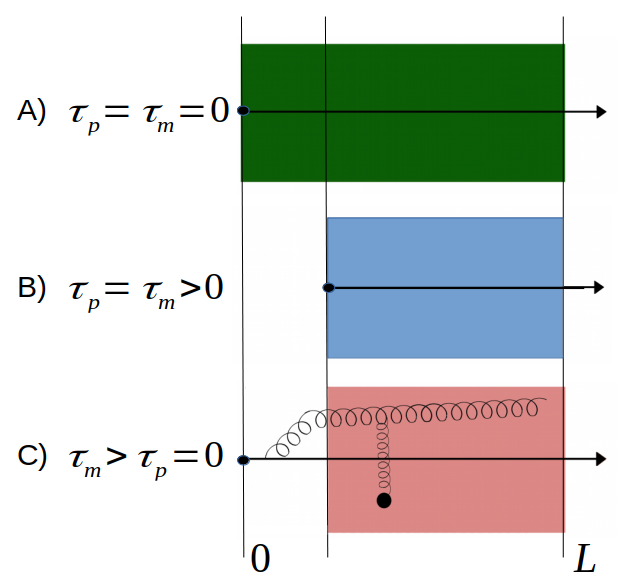}
\caption{Schematic representation of the three different scenarios analyzed: case A: $\tau_{\mathrm{p}}=\tau_{\mathrm{m}} =0$ (top panel); case B: $\tau_{\mathrm{p}} =\tau_{\mathrm{m}} > 0$ (middle panel); and case C: $\tau_{\mathrm{m}} >\tau_{\mathrm{p}}=0$ (bottom panel). The emitted gluon on the bottom panel illustrates the kind of contributions which are only accounted for in case C due to emissions occurring at times prior to $\tau _{\rm m}$ in addition to the emissions inside the medium which are included in all cases. }
\label{fig:picture_cases}
\end{figure}

\begin{enumerate}[label=\textbf{Case \Alph*:},align= left]
\item $\tau_{\rm p} =\tau_{\rm m}=0$. The emitter is produced at the same time as the medium is formed. In this scenario the hard parton propagates through a medium with longitudinal extension going from  0 to $L$. This corresponds to the standard setup of theoretical calculations of medium-induced radiation.
\item $\tau_{\rm p } =\tau_{\rm m } > 0$. The hard parton is produced at the same time as the medium is formed but this time is no longer zero. This case is formally equivalent to the previous one, but for a medium with shorter length given by $L -\tau_{\rm m}$. 
\item $\tau_{\rm m} > \tau_{\rm  p}=0$. The emitter is produced at $\tau_{\rm p} = 0$ 
propagating first through vacuum from $0$ to $\tau_{\rm m }$ and then throughout a medium of longitudinal extension going from $\tau_{\rm m}$ to $L$. This scenario corresponds to a medium  with length $L-\tau_{\rm m}$, but due to the emitter's propagation through vacuum for times prior to $\tau_{\mathrm{m}}$ takes into account extra medium-induced radiation with respect to case B arising from emissions happening at times prior to $\tau_{\rm m}$.
\end{enumerate}

In this manuscript, we first derive the medium-induced radiation spectrum for the general case where $\tau_{\rm m} \geq \tau_{\rm p}$  within the BDMPS-Z framework \cite{Baier:1996kr, Baier:1996sk,Zakharov:1996fv,Zakharov:1997uu}  which includes the resummation of multiple in-medium scatterings, and then isolate the contribution from allowing the initial hard parton to radiate before entering the medium. Then we compute, by means of this energy loss calculation, the charged hadron nuclear modification factor $R_{\rm AA}$ and the high-$p_T$ azimuthal asymmetry $v_2$ for the $20\text{-}30\%$  centrality class in $\sqrt{s_{\mathrm{NN}}}=2.76\,{\rm TeV}$ Pb-Pb collisions in a simplified set-up, thus quantifying the impact of this initial radiation on this set of jet quenching observables.

The paper is organized as follows: in section~\ref{sec:full} we derive the fully resummed energy spectrum and perform, following \cite{Andres:2020vxs,Andres:2020kfg}, its numerical evaluation   for the  three initial stage scenarios above described. In section~\ref{sec:GLV} we obtain and compare the first opacity limit of the BDMPS-Z spectrum \cite{Gyulassy:2000fs, Gyulassy:2000er,Wiedemann:2000za}, also known as GLV limit, within these scenarios. In section~\ref{sec:HO} we turn our attention to the  harmonic oscillator (HO) approximation, which by assuming a Gaussian profile for the transverse momentum transfers yields a (semi)-analytical expression for the in-medium energy spectrum. In section~\ref{sec:v2} we present the analysis on the simultaneous description of the single-inclusive hadron suppression and the harmonic coefficient $v_2$ in the hard sector both within the HO and fully resummed formalisms for the three early time scenarios considered. Finally, we summarize and conclude in section~\ref{sec:conclusions}.

\section{Fully resummed BDMPS-Z spectrum}
\label{sec:full}
We start by deriving the BDMPS-Z medium-induced radiation spectrum for generic $\tau_{\rm p}$ and $\tau_{\rm m }$. First we recognize that $\tau_{\rm p}$ enters as the lower limit of all the longitudinal integrations, since we are only concerned with what happens after the hard parton is created. We take as a starting point  the medium-induced $\vec{k}$-differential spectrum off a highly energetic parton given in eq.~(3.3) of \cite{Andres:2020vxs}, which after introducing $\tau_{\rm p}$ takes the following form\footnote{Throughout this manuscript bold font will be used for 2D vectors in the transverse plane with respect to the propagation of the leading parton. We adopt the shorthand $\int_{\vec{p}}= \int \mathrm{d}^2\vec{p}/(2\pi)^2$ for the transverse integrals in momentum space.} 
\beq
    \omega \frac{\mathrm{d}I^{\mathrm{Full}}}{\mathrm{d} \omega \mathrm{d}^2 \vec k}=
    \frac{ \alpha_{\rm s}C_{\rm R}}{ 2\pi^2 \omega} \operatorname{Re} 
    \int_{\tau_{\rm p}}^{\infty} \mathrm{d}s \, n(s) 
    \int_{\tau_{\rm p}}^s \mathrm{d}t 
    \int_{\vec p \vec q \vec l} 
    i \vec{p} \cdot\left(\frac{\vec l}{\vec l^2}-\frac{\vec q}{\vec q^2}\right) 
    \sigma(\vec l-\vec q ) 
   \widetilde{\cK}(s, \vec q ; t, \vec p) 
    \cP(L, \vec k ; s, \vec l)\,,
\label{eq:Full_spectrum}
%\eqref{eq:Full_spectrum}
\eeq
where the vacuum contribution has already been subtracted. $C_{\rm R}$ denotes the Casimir of the color representation  of the emitter, and $\omega$ and $\vec{k}$ are,  respectively, the energy and transverse momentum of the emitted soft gluon. The linear density of scattering centers is denoted by $n(s)$  and $\sigma(\vec q)$  is the dipole cross section which contains the specific details of the parton-medium interaction model through the elastic collision rate $V(\vec{q})$
\beq
\sigma(\vec{q}) = - V(\vec q) + (2\pi)^2\delta^{(2)}(\vec q)\int_{\vec l}V(\vec l)\, .
\label{eq:sigmaV}
\eeq
$\cP(L,\vec k;s,\vec l)$ and $\widetilde\cK (s,\vec q;t,\vec p )$  are, respectively,  the  transverse momentum broadening and emission kernel  in momentum space, which satisfy the following Schwinger-Dyson equations
\begin{align}
\cP(t',\vec k;t,\vec l) &= \left(2\pi\right)^2\delta^{(2)}(\vec k-\vec l )\,-\,\frac{1}{2} 
\int_t^{t'} \mathrm{d}s \,n(s) \int_{\vec k'} \, 
\sigma(\vec k'-\vec l) \cP(t',\vec k;s,\vec k')\: 
,\label{eq:PSD1}\\
\widetilde{\cK}(t',\vec q;t,\vec p ) 
&= \left(2\pi\right)^2\delta^{(2)}(\vec q-\vec p )\,e^{-i \frac{\vec p^2}{2\omega}(t'-t)}\nonumber\\
&\quad \qquad-\,\frac{1}2  
\int_t^{t'} \mathrm{d}s \,n(s) \int_{\vec k'}  \,\sigma(\vec q -\vec k')
e^{-i \frac{\vec q^2}{2\omega}(t'-s)} 
\widetilde{\cK}(s,\vec k';t,\vec p)\:.
\label{eq:KSD1}
\end{align}

The linear density $n(s)$ is different from zero only in a finite region corresponding to the longitudinal extension of the medium that goes from $\tau_{\rm m}$ to $L$. Weather $\tau_{\rm m} = \tau_{\rm p}$ or not will depend on the cases A, B and C outlined in section~\ref{sec:intro}. As such, the integration over $s$ in~\eqref{eq:Full_spectrum} goes then from $\tau_{\rm m}$ to $L$,
\beq
\omega \frac{\mathrm{d}I^{\mathrm{Full}}}{\mathrm{d} \omega \mathrm{d}^2 \vec k}=
\frac{ \alpha_{\rm s}C_{\rm R}}{ 2\pi^2 \omega} \operatorname{Re} 
\int_{\tau_{\rm m}}^L \mathrm{d}s \, n(s) 
\int_{\tau_{\rm p}}^s \mathrm{d}t 
\int_{\vec p \vec q \vec l} 
i \vec{p} \cdot\left(\frac{\vec l}{\vec l^2}-\frac{\vec q}{\vec q^2}\right)  \sigma(\vec l-\vec q ) 
 \widetilde{\cK}(s, \vec q ; t, \vec p) 
\cP(L, \vec k ; s, \vec l)\,.
\label{eq:Full_spectrum2}
%\eqref{eq:Full_spectrum2}
\eeq
By setting $\tau_{\rm p} = \tau_{\rm m} =0$, we obtain the standard scenario where the parent parton is produced at a time $0$  and propagates through a medium with longitudinal extension going from $0$ to $L$ (see top panel of figure~\ref{fig:picture_cases}). The numerical evaluation of this scenario, consisting of solving the differential equations satisfied by the broadening and the kernel, which can be obtained as time derivatives of eqs.~\eqref{eq:PSD1}~and~\eqref{eq:KSD1} respectively \cite{Zakharov:2004vm,Caron-Huot:2010qjx}, was explained in detail and performed for a static medium in \cite{Andres:2020vxs}.

In the more general case where the hard parton is produced before the formation of the medium ($\tau_{\rm m} \geq \tau_{\rm p}$) we can isolate the contribution from the initial radiation by splitting the $t$-integration at $\tau_{\rm m}$. Eq.~\eqref{eq:Full_spectrum2} then takes the form 
\beq
    \begin{aligned}
        \omega\frac{\mathrm{d}I^{\mathrm{Full}}}
        {\mathrm{d} \omega \mathrm{d}^2 \vec k} =
        \,&\frac{ \alpha_{\rm s}C_{\rm R}}{2 \pi^2 \omega} \operatorname{Re}
        \int_{\tau_{\rm m}}^L \mathrm{d}s \, n(s) 
        \int_{\tau_{\rm p}}^{\tau_{\mathrm m}} \mathrm{d}t 
        \int_{\vec p  \vec q \vec l} 
        i \vec{p} \cdot\left(\frac{\vec l}{\vec l^2}-\frac{\vec q}{\vec q^2}\right)  \sigma(\vec l-\vec q )
       \widetilde{\cK}(s,\vec q; t,\vec p) \,
        \cP(L,\vec k; s,\vec l) 
        \\
        &+\frac{ \alpha_{\rm s}C_{\rm R}}{2 \pi^2 \omega} \operatorname{Re}
        \int_{\tau_{ \rm m}}^L \mathrm{d}s \,n(s) 
        \int_{\tau_{ \rm m}}^s \mathrm{d}t
        \int_{\vec p \vec q \vec l} 
        i \vec{p} \cdot\left(\frac{\vec l}{\vec l^2}-\frac{\vec q}{\vec q^2}\right)  \sigma(\vec l-\vec q)
        \widetilde{\cK}(s, \vec q ; t, \vec{p}) \,
        \cP(L, \vec k ; s, \vec l)\,.
    \end{aligned}
%\eqref{eq:Full_spectrum3}
\label{eq:Full_spectrum3}
\eeq
The first term in \eqref{eq:Full_spectrum3} can be further simplified employing the convolution property of the kernel
\beq
    \widetilde{\cK}(s, \vec q ; t, \vec p) = \int_{\vec k_1}
    \widetilde{\cK}(s, \vec{q} ; \tau_{\mathrm m}, \vec k _1)\,
    \widetilde{\cK}_0(\tau_{\mathrm m}, \vec k_1 ; t, \vec{p}) = 
    \widetilde{\cK}(s, \vec q ; \tau_{\mathrm m}, \vec p)\,
    e^{-i\frac{ \vec p^2}{2\omega}(\tau_{\mathrm m}-t) }\,,
    \label{eq:Kconv}
\eeq
where we have used that for times prior to $\tau_{\mathrm m}$ the kernel is the vacuum one $\widetilde{\cK}_0$, which, according to eq.~(\ref{eq:KSD1}), is given by
\beq
\widetilde{\cK}_0(\tau_{\mathrm m}, \vec k_1 ; t, \vec p)= 
(2\pi)^2 \delta^2(\vec k_1-\vec p) e^{-i\frac{ \vec p^2}{2\omega}(\tau_{\mathrm m}-t) } \,.
\label{eq:KernelVac}
\eeq
Plugging (\ref{eq:Kconv}) into the first line of (\ref{eq:Full_spectrum3}) and performing  the integral over $t$, we obtain
\beq
\begin{aligned}
\omega \frac{\mathrm{d}I^{\mathrm{Full}}}
{\mathrm{d} \omega \mathrm{d}^2 \vec k} &=
\frac{\alpha_{\rm s}C_{\rm R}}{\pi^2 } 
\operatorname{Re}
\int_{\tau_{\mathrm m}}^L \mathrm{d} s \, n(s)
\int_{\vec p \vec q \vec l} \frac{\vec p}{\vec p^2} 
\cdot\left(\frac{\vec l}{\vec l^2}-\frac{\vec q}{\vec q^2}\right) 
\left(1 - e^{-i\frac{\vec p^2}{2\omega}(\tau_{\mathrm m}-\tau_{\mathrm p}) } \right) 
\sigma(\vec l-\vec q) \\ 
& \qquad \quad\times
\widetilde{\cK}(s, \vec q ; \tau_{\mathrm m}, \vec p)\, 
\cP(L, \vec k ; s, \vec l)
\\
&\quad + 
\frac{\alpha_{\rm s}C_{\rm R}}{2\pi^2 \omega} \operatorname{Re} 
\int_{\tau_{\mathrm m}}^L \mathrm{d}s \, n(s) 
\int_{\tau_{\mathrm m}}^{s} \mathrm{d}t 
\int_{\vec p \vec q \vec l} 
i \vec{p} \cdot\left(\frac{\vec l}{\vec l^2}-\frac{\vec q}{\vec q^2}\right) 
\sigma(\vec l-\vec q) \,
\widetilde{\cK}(s, \vec q ; t, \vec p)\, 
\cP(L, \vec k ; s, \vec l)\,.
\label{eq:Full_spectrum4}
%\eqref{eq:Full_spectrum4}
\end{aligned}
\eeq
The second term in \eqref{eq:Full_spectrum4} corresponds to the spectrum in the case where the hard parton is created at the same time as the starting point of the medium ($\tau_{\rm p} = \tau_{\rm m}$) while the first term accounts for the additional radiation due to the initial propagation through vacuum. This first term is clearly zero when $\tau_{\rm p} = \tau_{\rm m}$.

Throughout this manuscript, we  focus on the energy distribution, which can be obtained by integrating the $\vec k$-differential spectrum over transverse momentum\footnote{We use the notation $k=|\vec k|$ for the magnitude of 2D vectors in the transverse plane.} $\vec k$
\beq
\omega \frac{\mathrm{d}I^\mathrm{Full}}{\mathrm{d}\omega} =
\int_0^\omega \mathrm{d}k  \,k 
\int_0^{2\pi}  \mathrm{d} \theta_k\,
\omega \frac{\mathrm{d}I^{\mathrm{Full}}}
{\mathrm{d} \omega \mathrm{d}^2 \vec k}\,,
\label{eq:spec_rfin}
%\eqref{eq:spec_rfin}
\eeq
where we have imposed the kinematical constraint restricting the transverse momentum of the radiated gluon to be smaller than its energy $k \leq \omega$. The numerical evaluation of the energy distribution follows the procedure explained in ref.~\cite{Andres:2020vxs}, which consists of solving the differential equations satisfied by the broadening factor and the evolution kernel. Once the propagators have  been obtained, we can plug them into the expression of the spectrum and perform the integrations numerically. For further details, we refer the reader to~\cite{Andres:2020vxs}.

We also explore the situation where we  remove the kinematical condition by performing the $\vec k$-integration over all the transverse momentum phase space ($0 \leq k < \infty$), in which case the broadening factor is integrated out and no longer plays a role in the evaluation. It is worth noticing that  the derivation of this spectrum assumes the transverse momentum of the emitted gluon is much smaller than its energy ($k \ll \omega$). Hence, applying this limit does not correspond to any realistic physical situation leading to a divergent spectrum for small values of $\omega$. Contrarily, when a realistic kinematic constraint on the transverse momentum of the radiated gluon in imposed, the energy spectrum is depleted at low $\omega$. Performing  the $\vec k$-integration without the kinematical condition  in \eqref{eq:Full_spectrum4} and realizing that second term in brackets is zero in both terms, we obtain
\beq
\begin{aligned}
   \omega \frac{\mathrm{d}I^{\mathrm{Full}}}{\mathrm{d} \omega }=\,
    & 4 \alpha_{\rm s}C_{\rm R} \operatorname{Re} 
    \int_{\tau_{\mathrm m}}^L \mathrm{d}s \, n(s) 
    \int_{\vec p \vec q \vec l} 
     \frac{ \vec{p} \cdot\vec l}{\vec p^2\,\vec l^2}\,
     \left (1- e^{-i \frac{\vec p^2}{2\omega}(\tau_{\mathrm m}-\tau_{\mathrm p})} \right)
    \sigma(\vec l-\vec q ) \,
    \widetilde{\cK}(s, \vec q ; t, \vec p)  \\ 
     &+
     \frac{2 \alpha_{\rm s}C_{\rm R}}{\omega} \operatorname{Re} 
    \int_{\tau_{\mathrm m}}^L \mathrm{d}s \, n(s) 
    \int_{\tau_{\mathrm m}}^s \mathrm{d}t 
    \int_{\vec p \vec q \vec l} 
     i \,\frac{ \vec{p} \cdot\vec l}{\vec l^2}\,
    \sigma(\vec l-\vec q ) \,
    \widetilde{\cK}(s, \vec q ; t, \vec p)\,.
\label{eq:CasecFull_rinf}
%\eqref{eq:CasecFull_rinf}
\end{aligned}
\eeq
In the particular case where we set $\tau_{\mathrm p}=\tau_{\mathrm m}$, the energy spectrum can be written as
\beq
   \omega \frac{\mathrm{d}I^{\mathrm{Full}}}{\mathrm{d} \omega }=
     \frac{2 \alpha_{\rm s}C_{\rm R}}{\omega} \operatorname{Re} 
    \int_{\tau_{\mathrm m}}^L \mathrm{d}s \, n(s) 
    \int_{\tau_{\mathrm m}}^s \mathrm{d}t 
    \int_{\vec p \vec q \vec l} 
     i \,\frac{ \vec{p} \cdot\vec l}{\vec l^2}\,
    \sigma(\vec l-\vec q ) \,
    \widetilde{\cK}(s, \vec q ; t, \vec p)\,.
\label{eq:CasecFull_rinf_AB}
\eeq

Throughout this paper we will evaluate the energy distributions (with and without the kinematic constraint) 
using the collision rate $V$ of a Yukawa  elastic parton-medium scattering, also known as Gyulassy-Wang model \cite{Gyulassy:1993hr}, given by
\beq
    V(\vec q) =
    \frac{8\pi \mu^2}{(\vec q^2 + \mu^2)^2}\,,
\label{eq:yukawa}
%\eqref{eq:yukawa}
\eeq
where $\mu^2$ is the screening mass.

Let us now restrict to a static scenario where the medium is modeled as a ``brick'' with constant screening mass $\mu^2(s)=\mu^2$ for $s \,\in\, [\tau_{\rm m},L]$, and constant linear density of scattering centers $n(s) = n_0 $ for $s \,\in\, [\tau_{\rm m},L]$. In this brick model, and given that we will only analyze in this manuscript cases where $\tau_{\mathrm p}$ is either  $\tau_{\rm p}=0$ or  $\tau_{\rm p} =\tau_{\rm m}$, the only medium parameters entering the in-medium emission spectra are: $n_0$, $L$, $\mu^2$ and $\tau_{\rm m}$. We  will use instead the following
\beq
 n_0L\,, \quad
 \frac{\tau_{\mathrm m}}{L}\,, \quad
 \bar{\omega}_{\rm c} \equiv \frac{1}{2}\mu^2L\,, \quad
\bar R \equiv \bar{\omega}_{\rm c}L \,,
\label{eq:variables_full}
%\eqref{eq:variables_full}
\eeq
where the dimensionless spectrum $\omega dI^{\mathrm{Full}}/d\omega$ is a function of $\omega/\bar\omega_{\rm c}$ with no additional dependence on either $\omega$ or $\omega_{\rm c}$ separately.  Taking the limit $\bar R \rightarrow \infty$ with fixed $\bar \omega_{\rm c}$ is equivalent to extend the integration over the transverse momentum of the radiated gluon up to infinity.

\begin{figure}
\centering
\includegraphics[width=\linewidth]{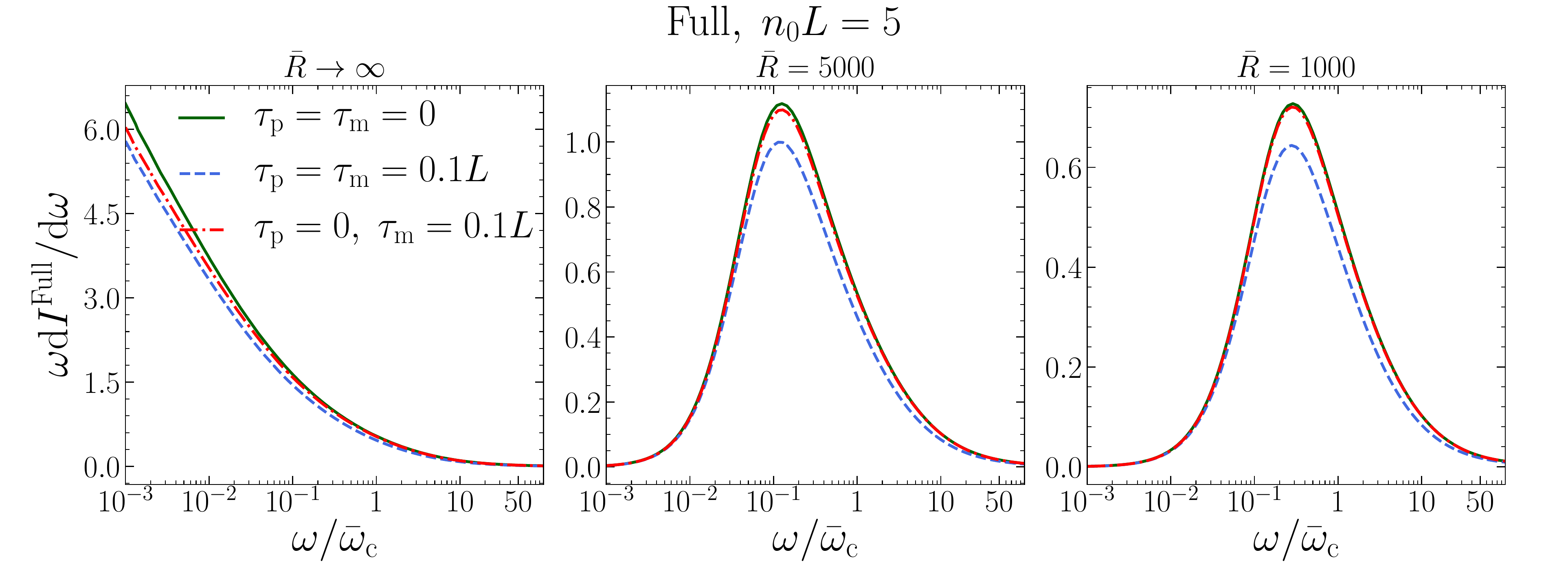}
\caption{Fully resummed medium-induced energy spectrum for $\tau_{\rm p} = \tau_{\rm m} = 0$ (green solid lines);  $\tau_{\rm p} = \tau_{\rm m} =0.1\,L $ (blue dashed lines); and $\tau_{\rm p} = 0$ and $\tau_{\rm m} =0.1\,L$  (red dash-dotted lines) as function of  $\omega/\bar \omega_{\rm c}$  for $n_0L=5$. The different panels correspond to different  sets of $\bar R $: $\bar R \rightarrow\infty$ (left), $\bar R =5000$ (central), and $\bar R =1000$ (right).}
\label{fig:full_t0L0_1}
\end{figure}

\begin{figure}
\centering
\includegraphics[width=\linewidth]{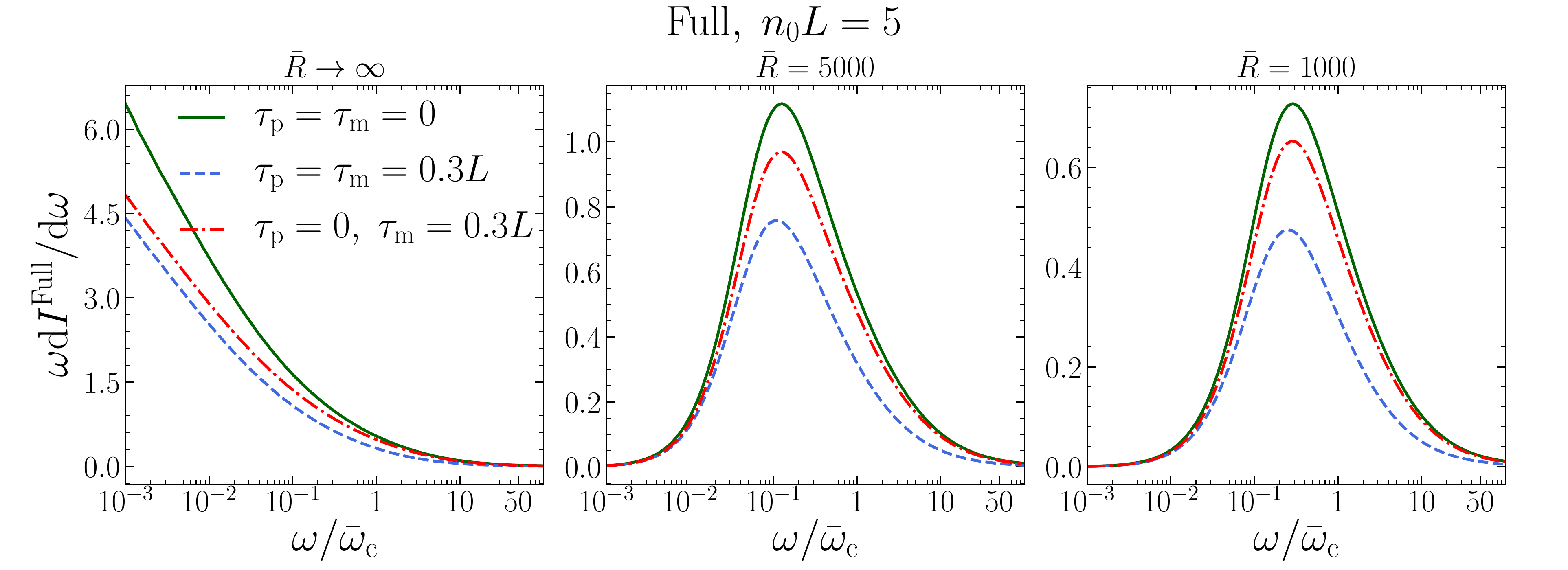}
\caption{Fully resummed medium-induced energy spectrum for the three scenarios:  $\tau_{\rm p} = \tau_{\rm m} = 0$ (green solid lines); $\tau_{\mathrm p} = \tau_{\rm m} =0.3\,L $ (blue dashed lines); and $\tau_{\rm p} = 0$ and $\tau_{\rm m} =0.3\,L$  (red dash-dotted lines) as function of  $\omega/\bar \omega_{\rm c}$  for $n_0L=5$. The different panels correspond to different  sets of $\bar{R}$: $\bar R\rightarrow \infty$ (left), $\bar R=5000$ (central), and $\bar R=1000$ (right).}
\label{fig:full_t0L0_3}
\end{figure}

We now present the numerical evaluation of the fully resummed energy distribution in the brick for three different initial stages scenarios depicted in section~\ref{sec:intro}. For illustration purposes, we will always consider in this manuscript  the case where the emitter is a quark, thus taking  $C_{\rm R} = C_{\rm F}= 4/3$, and we fix the strong coupling to $\alpha_s=0.3$. These energy spectra are shown in figures~\ref{fig:full_t0L0_1}~and~\ref{fig:full_t0L0_3} as a function of $\omega/\bar \omega_{\rm c} \equiv 2\omega/(\mu^2L)$ for a brick of $n_0L = 5$. In both figures, the three scenarios considered are represented as follows: the green solid lines correspond to $\tau_{\mathrm p} = \tau_{\rm m}=0$, the blue dashed lines to $\tau_{\rm p} = \tau_{\rm m} > 0$, and the red dash-dotted lines to $\tau_{\rm m} > \tau_{\rm p}=0$. We have set $\tau_{\rm m}/L = 0.1$ in  figure~\ref{fig:full_t0L0_1} and $\tau_{\mathrm m}/L = 0.3$ in figure~\ref{fig:full_t0L0_3}, except for the green curves in which $\tau_{\rm m}=0$ for both figures. We show on the left panel of both figures the energy spectra for the three early time cases without the kinematic constraint ($\bar R \rightarrow \infty$), while the curves on the central and right panels are the corresponding energy distributions with the kinematic constraint $ k \leq \omega $ imposed for two different values of the parameter $\bar R$: $\bar R$ = 5000 (central panel) and $\bar R = 1000$ (right panel).\footnote{The values of $\bar R$ are chosen sufficiently separated to show how the spectrum changes with this parameter while making sure that the peak is not in the high-energy tail.}

First, it is clear that the case where $\tau_{\rm p} = \tau_{\rm m} > 0$, is formally equivalent to the case where $\tau_{\rm p} = \tau_{\rm m} = 0$ for a medium with shorter longitudinal extension ($L- \tau_{\rm m}$ with $\tau_{\rm m} \neq 0$ instead of $L$). This is reflected in the fact that for a given set of medium parameters the curve corresponding to $\tau_{\rm p} = \tau_{\rm m} > 0$  (blue dashed lines) is always smaller than the one corresponding to $\tau_{\rm p} = \tau_{\rm m} = 0$  (green solid lines) with the difference growing with $\tau_{\rm m}$. This can be clearly appreciated in figures~\ref{fig:full_t0L0_1} and \ref{fig:full_t0L0_3}.

In the case where the additional initial radiation in vacuum is included ($\tau_{\rm m} > \tau_{\rm p}=0$), we can see that the curves are somewhere in between the two other cases. In figure~\ref{fig:full_t0L0_1}, where $\tau_{\rm m}/L$ is still small, the differences with the case where the medium starts at $\tau_{\rm m}=0$ (green solid lines) are only noticeable for very small values of $\omega/\bar\omega_{\rm c}$ (see left panel), while in figure~\ref{fig:full_t0L0_3} those differences grow and the two respective curves agree only in the high-energy tail. The fact that the emission of gluons with higher energies seems to be unaffected by removing part of the medium in the initial propagation can be explained in terms of formation times. Harder gluons require a larger transverse momentum transfer to decorrelate from the parent parton and therefore their formation time is longer. Those gluons with a formation time longer than $\tau_{\rm m}$ can be emitted before the medium starts and still interact with the medium before being completely decorrelated from the parent parton, while those with shorter formation times can only contribute to the medium-induced part of the spectrum if they are emitted after the medium is formed. In sections \ref{sec:GLV} and \ref{sec:HO}, where we use the single scattering (GLV) and the harmonic approximations respectively we will see how this can be seen in an analytical setting where the relevant energy scales are explicit.

%%%%%%%%%%%%%%%%%%%%%%%%%%%%%%%%%%%%%%%%%%%%%%%%%%%%%%%%%%%%%%%%%
%%%%%%%%%%%%%%%%%%%%%%%%%%%%%%%%%%%%%%%%%%%%%%%%%%%%%%%%%%%%%%%%%
%% GLV
\section{First opacity (GLV) limit}
\label{sec:GLV}
The first opacity limit of the BDMPS-Z spectrum, which we refer to as GLV limit, is the single scattering contribution to the fully resummed spectrum (see for instance \cite{Wiedemann:2000za, Andres:2020vxs}). In order to  obtain this limit, we take as starting point eq.~\eqref{eq:Full_spectrum2}, where one of the scatterings is already explicit in the form of the factor $\sigma$, and replace $\cP$ and $\widetilde\cK$ with their vacuum versions given by
\beq
\cP \left(t', \vec{k} ; t, \vec l\right)
 =(2 \pi)^2 \delta^{(2)}(\vec k-\vec l)\,,
\label{eq:BroadVac}
\eeq
and \eqref{eq:KernelVac}, respectively. Following this approach and  integrating over $t$ in \eqref{eq:Full_spectrum2}, we get that the GLV  $\vec k$-differential spectrum can be written as
\beq
\omega 
  \frac{\mathrm{d}I^{\mathrm{GLV}}}{\mathrm{d}\omega \mathrm{d}^2\vec k}= 
\frac{ \alpha_{\rm s}C_{\rm R}}{\pi^2}
\operatorname{Re} 
\int_{\tau_{\rm m}}^L \mathrm{d}s \,n(s)
\int_{\vec p}
\,\frac{1}{\vec p^2}\left(\frac{\vec p\cdot\vec k}{\vec k^2}-1\right)  
\left(1- e^{-i \frac{\vec p^2}{2 \omega}\left(s-\tau_{\mathrm p}\right)} \right) \sigma(\vec k-\vec p) \,.
\label{eq:GLV_diffkt}
%\eqref{eq:GLV_diffkt}
\eeq
From this result, it is easy to see that the spectrum in the general case, where the parton is produced before entering the medium ($\tau_{\rm m} > \tau_{\rm p}$, see lower panel of figure~\ref{fig:picture_cases}), can be expressed as two separate contributions from partons created inside the medium, each with different length (see top panel of figure~\ref{fig:picture_cases}),
\beq
\begin{aligned}
 \left.\omega 
    \frac{\mathrm{d}I^{\mathrm{GLV}}}
    {\mathrm{d} \omega \mathrm{d}^2 \vec k}\right|_{\tau_{\rm m} >\tau_{\rm p}} =
    \left.\omega 
    \frac{\mathrm{d}I^{\mathrm{GLV}}}
    {\mathrm{d} \omega \mathrm{d}^2 \vec k}\right|_{\tau_{\rm m} =\tau_{\rm p}} - \,
    \frac{\alpha_{\rm s}C_{\rm R}}{\pi^2 } 
    \operatorname{Re} 
    \int_{\tau_{\rm p}}^{\tau_{\rm m}} & \mathrm{d}s\, n(s) 
    \int_{\vec p} 
    \frac{1}{\vec p^2}\left(\frac{\vec p\cdot\vec k}{\vec k^2}-1\right)\\
    &\times \left(1 - e^{-i\frac{\vec p^2}{2\omega}(s-\tau_{\rm p})} \right) 
    \, \sigma(\vec k-\vec p)\,.
  \label{eq:GLV_diffkt_casea}
\end{aligned}
\eeq
The second term in the r.h.s.~is clearly the spectrum for a hard parton propagating in a medium from $\tau_{\rm p}$ to $\tau_{\rm m}$. The difference between the initial propagation being either in vacuum or in medium is precisely the amount of radiation one gets from an in-medium propagation during that initial slab. This result is specific to the GLV limit and does not hold for the fully resummed result of the previous section. In that case one cannot separate between emission interacting with the medium before or after $\tau_{\rm m}$ given that the multiple scatterings can occur anywhere. It is worth noticing  that the GLV limit has been showed to correctly describe the high-energy tails of the fully resummed spectrum \cite{Andres:2020vxs, Andres:2020kfg}, and thus this relation is expected to also hold for the full BDMPS-Z spectrum at sufficiently high $\omega$.

Now, we move to the energy distribution. As long as we do not perform the $s$ integration in \eqref{eq:GLV_diffkt} we can continue the evaluation of all the other integrals in the usual way. Using the definition of $\sigma$ in \eqref{eq:sigmaV} and the Yukawa potential of \eqref{eq:yukawa} one can perform explicitly the integral over $\vec k$ and the angular part of the $\vec p$ integral, yielding
\beq
\begin{aligned}
\omega 
  \frac{\mathrm{d}I^{\mathrm{GLV}}}{\mathrm{d}\omega }= 
 \frac{4\alpha_{\rm s}C_{\rm R}}{\pi}
\operatorname{Re} 
\int_{\tau_{\rm m}}^L \mathrm{d}s \,n(s)
\int_0^\infty & \frac{\mathrm{d} p}{p}
\left[
\frac{ \mu^2 }{\mu^2 + p^2 } -
\frac{ \mu^2 }{ \sqrt{(\mu^2 +  p^2+ \omega^2)^2 -4 \,\omega^2 \, p^2}}
\right]\\
&\times
 \left(1- e^{-i \frac{ p^2}{2 \omega}
 \left (s-\tau_{\rm p} \right)} \right)\,.
\label{eq:GLV_rfin}
%\eqref{eq:GLV_rfin}
\end{aligned}
\eeq
If the kinematical constraint is removed and the $k$ integration is allowed to go up to infinity the second term in the square brackets above disappears, thus getting
\beq
\omega \frac{\mathrm{d}I^{\mathrm{GLV}}}{\mathrm{d}\omega }
= \frac{4\alpha_{\rm s}C_{\rm R}}{\pi}
\operatorname{Re} 
\int_{\tau_{\rm m}}^L \mathrm{d}s \,n(s)
\int_0^\infty  \frac{\mathrm{d} p}{p}
\frac{ \mu^2 }{\mu^2 + p^2 } 
 \left(1- e^{-i \frac{ p^2}{2 \omega}(s-\tau_{\mathrm p})} \right)\,.
\label{eq:GLV_rinf}
%\eqref{eq:GLV_rinf}
\eeq
For both cases of the integrated energy spectrum (with and without kinematical constraint), the relation between the spectrum with initial vacuum propagation and the case with in-medium production explained after eq.~\eqref{eq:GLV_diffkt_casea} still holds.

\subsection{GLV spectrum in the brick}
\label{subsec:GLV_brick}
Let us now restrict to a static scenario where the medium is modeled as a ``brick'' with constant  linear density of scattering centers $n(s) = n_0$   for $s \in [ \tau_{\rm{m}},L]$, and constant screening mass $\mu(s) = \mu_0$  for $s \in [ \tau_{\rm{m}},L]$. Integrating over $s$ in \eqref{eq:GLV_diffkt},  we obtain that the GLV $\vec k$-differential spectrum in the brick is given by 
\beq
\begin{aligned}
\omega 
\frac{\mathrm{d}I^{\mathrm{GLV}}}
{\mathrm{d} \omega \mathrm{d}^2\vec k}
=\frac{ \alpha_{\rm s}C_{\rm R} \,n_0}{ \pi^2 }   
\int_{\vec p} \frac{1}{\vec p^2} \left(\frac{\vec p\cdot\vec{k}}{\vec k^2}
-1\right) 
&\sigma(\vec k-\vec p) 
\Bigg \{ L - \tau_{\rm m}  \\ 
& 
-\frac{2 \omega}{\vec p^2}\left[
\sin \left(\frac{\vec p^2 (L-\tau_p)}{2\omega}\right)   -
\sin \left(\frac{\vec p^2 (\tau_{\mathrm m}-\tau_ {\rm p})}{2\omega}\right) \right]
\Bigg\}\,.
\label{eq:GLVdiffkt_brick}
\end{aligned}
\eeq

For the energy spectrum, we first  look at the expressions where the kinematic constraint $k \leq \omega$ is imposed. Integrating over $s$ in \eqref{eq:GLV_rfin} we get
\begin{align}
\omega\frac{\mathrm{d}I^{\mathrm{GLV}}}{\mathrm{d}\omega }= 
\frac{4\alpha_{\rm s}C_{\rm R}n_0}{\pi}
\int_0^\infty & \frac{\mathrm{d} p}{p}
\left[
\frac{ \mu^2 }{\mu^2 + p^2 } -
\frac{ \mu^2 }{ \sqrt{(\mu^2 +  p^2+ \omega^2)^2 -4\omega^2p^2}}
\right]\nonumber\\
&\times
 \left(L-\tau_{\rm m}- \frac{2\omega}{p^2}\left[\sin\left(\frac{p^2(L-\tau_{\rm p})}{2 \omega}\right)-\sin\left(\frac{p^2(\tau_{\rm m}-\tau_{\rm p})}{2 \omega}\right)\right]\right)\,.
\label{eq:GLV_rfinbr}
\end{align}
In order to reduce the total number of parameters on which the result depends, we rescale the momentum variable $p \to p\sqrt{2\omega/L}$ obtaining
\begin{align}
\omega\frac{\mathrm{d}I^{\mathrm{GLV}}}{\mathrm{d}\omega }= 
\frac{4\alpha_{\rm s}C_{\rm R}n_0L}{\pi}
\int_0^\infty & \frac{\mathrm{d} p}{p}
\left[
\frac{1}{1 + xp^2} -
\frac{1}{\sqrt{(1 + xp^2+ x^2\bar R/2)^2 -2x^3\bar R p^2}}
\right]\nonumber\\
&\times\left(1-\frac{\tau_{\rm m}}{L}- \frac{1}{p^2}\left[\sin\left(p^2(1-\frac{\tau_{\rm p}}{
L})\right)-\sin\left(\frac{p^2}{L}(\tau_{\rm m}-\tau_{\rm p})\right)\right]\right)\,,
\label{eq:GLV_rfin_br_rescaled}
\end{align}
where we have defined $x\equiv \omega/\bar\omega_{\rm c}$, with $\bar\omega_{\rm c}$ and $\bar R$ given  in \eqref{eq:variables_full}. As previously,  $\tau_{\rm p}$ is set to either $\tau_{\rm p}=0$ or $\tau_{\rm p}=\tau_{\rm m} $ and thus  the GLV energy spectrum \eqref{eq:GLV_rfin_br_rescaled}  is a function of $n_0L$, $x$, $\bar R$ and $\tau_{\rm m}/L$. 

\begin{figure}
\centering
\includegraphics[width=\linewidth]{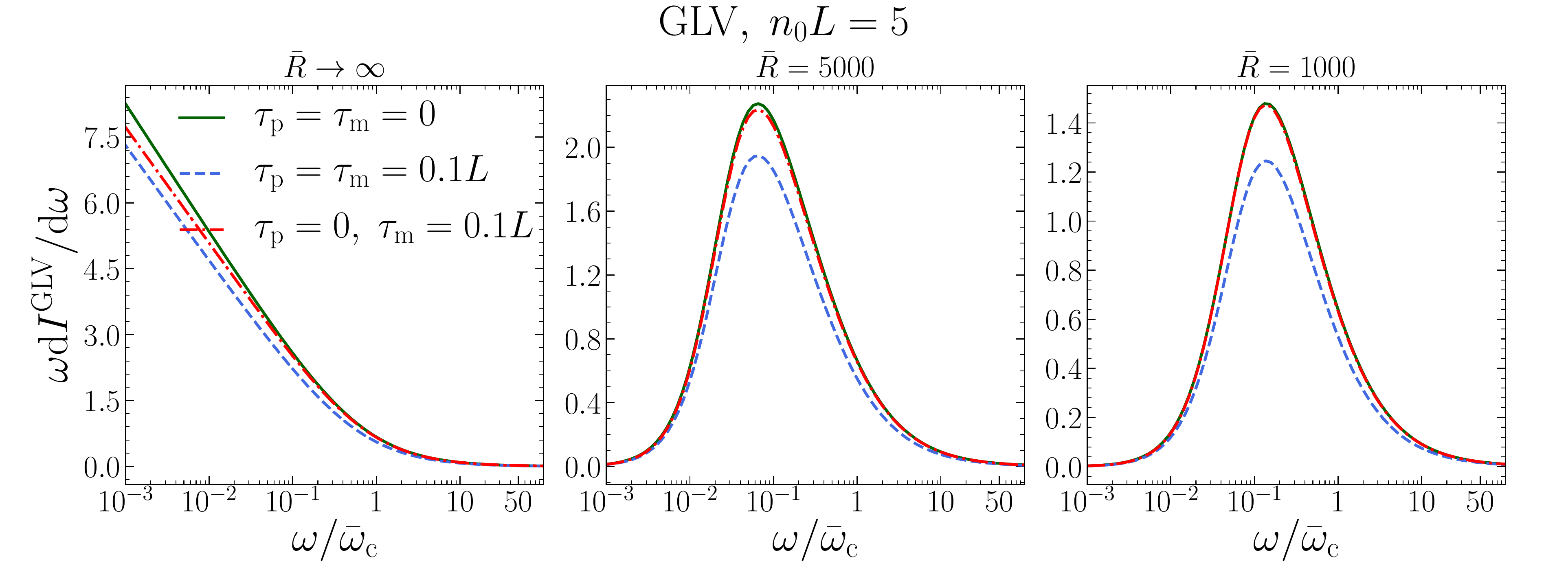}
\caption{GLV medium-induced energy spectrum for $\tau_{\rm p} = \tau_{\rm m} = 0$ (green solid lines);  $\tau_{\rm p} = \tau_{\rm m} =0.1\,L $ (blue dashed lines); and $\tau_{\rm p} = 0$ and $\tau_{\rm m} =0.1\,L$ (red dash-dotted lines) as function of $\omega/\bar \omega_{\rm c}$  for $n_0L=5$. The different panels correspond to different sets of $\bar R$: $\bar R\rightarrow\infty$ (left), $\bar R=5000$ (central), and $\bar R=1000$ (right).}
\label{fig:GLV_t0L0_1}
\end{figure}

Finally we obtain the expression of the GLV energy spectrum when the kinematic cut-off is removed ($0 \leq k <\infty$), in which case the second term in the brackets in the first line of \eqref{eq:GLV_rfinbr} vanishes, thus yielding
\begin{align}
\omega\frac{\mathrm{d}I^{\mathrm{GLV}}}{\mathrm{d}\omega }= 
\frac{4\alpha_{\rm s}C_{\rm R}n_0}{\pi}
\int_0^\infty & \frac{\mathrm{d} p}{p}
\frac{ \mu^2 }{\mu^2 + p^2 } 
 \left(L-\tau_{\rm m}- \frac{2\omega}{p^2}\left[\sin\left(\frac{p^2(L-\tau_{\rm p})}{2 \omega}\right)-\sin\left(\frac{p^2(\tau_{\rm m}-\tau_{\rm p})}{2 \omega}\right)\right]\right)\,.
\label{eq:GLV_rinfbr}
\end{align}
Setting $\tau_{\rm p} = 0$ it is straightforward to see that the result above is indeed the difference between the GLV spectra for a parton produced inside the medium traversing bricks of length $L$ and $\tau_{\rm m}$ respectively. Separating the two terms and rescaling the momentum variables accordingly, we can then write
\begin{align}
\omega\frac{\mathrm{d}I^{\mathrm{GLV}}}
{\mathrm{d}  \omega} = 
\frac{4\alpha_{\rm s}C_{\rm R}}{\pi}&\left(  
n_0L\int_0^\infty \mathrm{d}p \,
\frac{p^2 - \sin p^2}{p^3}\frac{1}{ 1+ \frac{\omega}{\bar \omega_{\rm c}}p^2} - n_0\tau_{\rm m}\int_0^\infty \mathrm{d}p \,
\frac{p^2 - \sin p^2}{p^3}\frac{1}{ 1+ \frac{\omega}{\bar \omega_{\rm m}}p^2}\right)\,,
\label{eq:GLV_rinf_brick_resc}
\end{align}
with $\bar\omega_{\rm m} \equiv \mu^2\tau_{\rm m}/2 = \bar\omega_{\rm c}\,\tau_{\rm m}/L$. The second term above becomes small when $\omega\gg\bar\omega_{\rm m}$ thus showing explicitly that for very large radiated gluon energies the initial part of the propagation does not affect the spectrum yielding identical spectra for emitters created inside or outside the medium. It can also be seen that the spectrum \eqref{eq:GLV_rinf_brick_resc} is a function of $n_0L$, $\tau_{\rm m}/L$, and $\omega/\bar\omega_{\rm c}$ only.
%%%%%%%%%%%%%%%%%%%%%%%%%%%%%%%%%%%

\begin{figure}
\centering
\includegraphics[width=\linewidth]{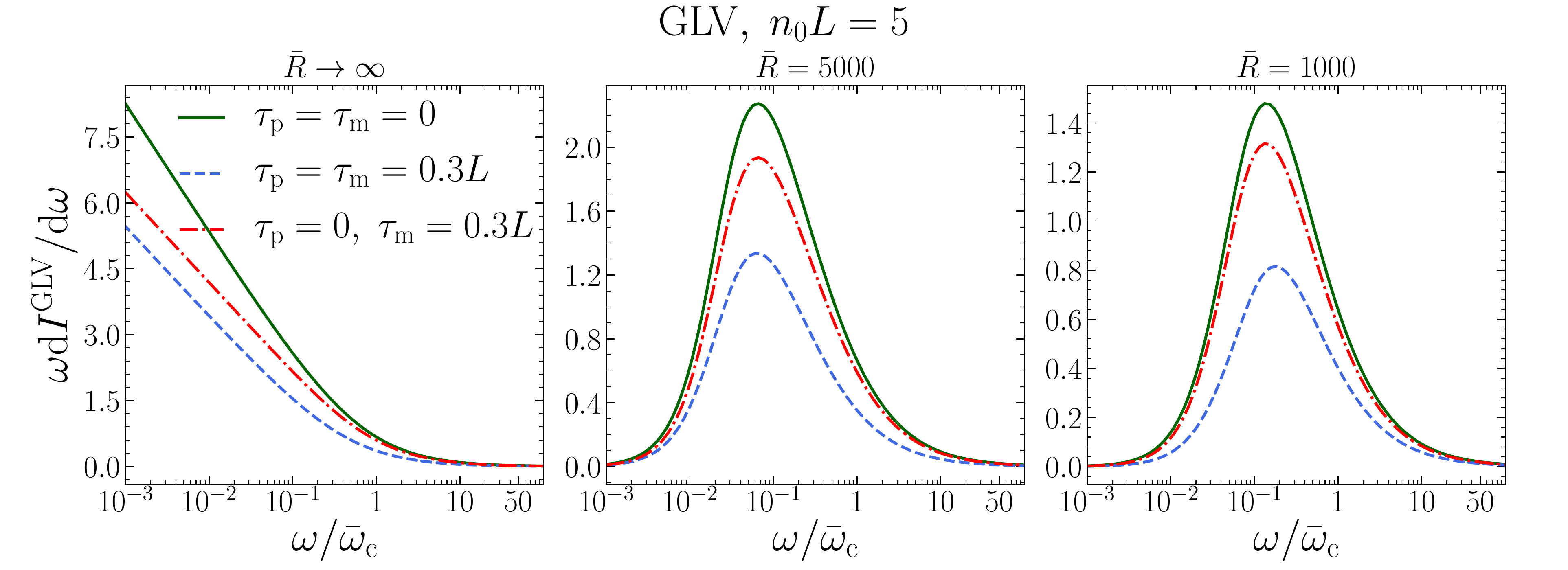}
\caption{GLV medium-induced energy spectrum for the three scenarios: $\tau_{\rm p} = \tau_{\rm m} = 0$ (green solid lines); $\tau_{\rm p} = \tau_{\rm m} = 0.3 \,L $ (blue dashed lines); and  $\tau_{\rm p} = 0$ and  $\tau_{\rm m} =0.3\,L$  (red dash-dotted lines) as function of $\omega/\bar \omega_{\rm c}$  for $n_0L=5$. The different panels correspond to different sets of $\bar R$: $\bar R \rightarrow\infty$ (left), $\bar R=5000$ (central), and $\bar R=1000$ (right).}
\label{fig:GLV_t0L0_3}
\end{figure}

We present in figures~\ref{fig:GLV_t0L0_1}~and~\ref{fig:GLV_t0L0_3} the GLV in-medium energy distribution off a high energy quark as a function of $\omega/\bar \omega_{\rm c} \equiv 2\omega/(\mu^2L)$ for the three scenarios  considered in this manuscript.  In both figures we have used the same brick parameters ($n_0L=5$) and kept the same line styles as in section~\ref{sec:full}:  green solid lines for $\tau_{\mathrm p} = \tau_{\rm m}=0$; blue dashed lines for  $\tau_{\rm p} = \tau_{\rm m} > 0$; and red dash-dotted lines for  $\tau_{\rm m} > \tau_{\rm p}=0$.  On the left panel of both figures we show the energy spectrum without the kinematic constraint ($ \bar R \rightarrow \infty$), while  the central and right panels show the energy distribution where the kinematic condition $k \leq \omega $ has been imposed  with two different values of the parameter $\bar R$: $\bar R=5000$ (central panel)  and $\bar R=1000$ (right panel). Figures~\ref{fig:GLV_t0L0_1}~and~\ref{fig:GLV_t0L0_3} differ in the value of $\tau_{\rm m}$ used for the curves where it is not zero (blue dashed and red dash-dotted), with $\tau_{\rm m}/L = 0.1$ in figure~\ref{fig:GLV_t0L0_1} and $\tau_{\rm m}/L = 0.3$ in figure~\ref{fig:GLV_t0L0_3}.

The results shown in figures~\ref{fig:GLV_t0L0_1}~and~\ref{fig:GLV_t0L0_3} agree with all the observations made in the previous section: the case with propagation in vacuum before entering the medium ($\tau_{\rm m} > \tau_{\rm p}=0$, red dash-dotted lines) is somewhere in between the other two cases where the parton is created inside the medium with varying lengths. At high energies, the tail of this spectrum coincides with that of having the full propagation inside the medium ($\tau_{\rm p} = \tau_{\rm m}=0$, green solid lines) but differs at low values of the energy. The transition occurs, as anticipated in this section, when $\omega/\bar\omega_{\rm c} \sim \tau_{\rm m}/L$ (or equivalently $\omega\sim\bar\omega_{\rm m}$), which can be seen from comparing figures~\ref{fig:GLV_t0L0_1}~and~\ref{fig:GLV_t0L0_3}.

%%%%%%%%%%%%%%%%%%%%%%%%%%%%%%%%%%%%%%%%%%%%%%%%%%%%%%%%%%%%%%%%%
%%%%%%%%%%%%%%%%%%%%%%%%%%%%%%%%%%%%%%%%%%%%%%%%%%%%%%%%%%%%%%%%%
%% HO
\section{Harmonic Oscillator approximation}
\label{sec:HO}
The Harmonic Oscillator or multiple soft scattering approach relies on approximating the dipole cross section in coordinate  space $\sigma(\vec{r})$ by its leading logarithmic behavior 
\beq
n(s)\sigma(\vec r) \equiv 
n(s)\int_{\vec q }
e^{i\vec q\cdot\vec r}\, \sigma({\vec q})
\approx \frac{1}{2} \hat{q}(s)\,\vec r^2 + 
\mathcal{O}(\vec r^2 \ln \vec r^2) \,,
\label{eq:sigma_HO}
\eeq
where $\hat{q}$ is the transport coefficient that describes the average transverse momentum squared transferred from the medium to the emitter per unit path length.  Since the natural way of expressing the HO in-medium spectrum is in coordinate space, we start by Fourier transforming eq.~\eqref{eq:Full_spectrum2} to coordinate space 
\begin{equation}
\begin{aligned}
\omega 
\frac{\mathrm{d} I}
{\mathrm{d} \omega \mathrm{d}^2 \vec k}=
\frac{2 \alpha_{\rm s}C_{\rm R}}{(2 \pi)^3 \omega} 
\operatorname{Re}
&\int_{\tau_{\rm m}}^L \mathrm{d} s \, n(s)
\int_{\tau_{\rm p}}^s \mathrm{d} t 
\int \mathrm{d}^2\vec x \,
\mathrm{d}^2\vec y \,i
\left[\sigma(\vec x) - \sigma(\vec y) \right]
\\&\times
\frac{\vec x - \vec y}{(\vec x-\vec y)^2} 
\cdot\partial_{\vec r}
\cK \left.(s, \vec x ; t, \vec r)\right|_{\vec r=0}
e^{-\frac{1}{2}\int_s^L 
\mathrm{d}\xi n(\xi)\sigma(\vec y)}
e^{-i\vec k\cdot \vec y}\,,
\end{aligned}
\label{eq:kt_spectrum_xspace}
\end{equation}
where we have used
\beq
\cP(L, \vec k ; s, \vec l) =
\int \mathrm{d}^2 \vec y\,
e^{-i(\vec k-\vec l)\cdot\vec y} 
e^{-\frac{1}{2} \int_s^L
\mathrm{d}\xi n(\xi)\sigma(\vec y)}\,,
\label{eq:broad_confguration_space}
\eeq
and
\beq
\widetilde{\cK}(s, \vec q; t,\vec p) =
\int  \mathrm{d}^2\vec x\,
\mathrm{d}^2\vec r\,
e^{-i(\vec q\cdot \vec x -  \vec p \cdot \vec r)} 
\cK(s, \vec x ; t, \vec r)\,.
\label{eq:kernel_configuration_space}
\eeq
In the general case where the emitter is produced before the formation of the medium
($\tau_{\rm m} \geq \tau_{\rm p}$) we can isolate the contribution coming from its initial radiation by splitting the $t$-integration in \eqref{eq:kt_spectrum_xspace}  at $\tau_{\rm m}$
\beq
\begin{aligned}
\omega 
\frac{\mathrm{d} I}{\mathrm{d} \omega \mathrm{d}^2 \vec{k}}=\,
&\frac{2 \alpha_{\rm s}C_{\rm R}}{(2 \pi)^3 \omega} 
\operatorname{Re}
\int_{\tau_{\mathrm m}}^L \mathrm{d} s \, n(s)
\int_{\tau_{\mathrm p}}^{\tau_{\rm m}} \mathrm{d} t 
\int \mathrm{d}^2\vec x \,
\mathrm{d}^2 \vec y \,i
\left[\sigma(\vec x) - \sigma(\vec y) \right]
\\& \qquad \quad \times
\frac{\vec x - \vec y}
{(\vec x-\vec y )^2}  
\cdot\partial_{\vec r }
\cK \left.(s, \vec x ; t, \vec r)\right|_{\vec r=0}
e^{-\frac{1}{2}\int_s^L
\mathrm{d}\xi n(\xi)\sigma(\vec y)}
e^{-i\vec k\cdot\vec y}
\\& 
+\frac{2 \alpha_{\rm s}C_{\rm R}}{(2 \pi)^{3} \omega} 
\operatorname{Re}
\int_{\tau_{\rm m}}^L \mathrm{d} s \, n(s)
\int_{\tau_{\rm m}}^s \mathrm{d} t 
\int \mathrm{d}^2\vec x \,
\mathrm{d}^2 \vec y \,i
\left[\sigma(\vec x) - \sigma(\vec y) \right]
\\&\qquad  \quad \times 
\frac{\vec x - \vec y}{(\vec x-\vec y)^2} 
\cdot\partial_{\vec r}
\cK \left.(s, \vec x  ; t, \vec r)\right|_{\vec r=0}
e^{-\frac{1}{2}
\int_s^L \mathrm{d}\xi n(\xi)\sigma(\vec y)}
e^{-i\vec k\cdot\vec y}\,.
\end{aligned}
\label{eq:kt_spectrum_xspace2}
%\eqref{eq:kt_spectrum_xspace2}
\eeq
The first term in \eqref{eq:kt_spectrum_xspace2} can be further simplified using  the convolution property  of the kernel in coordinate space
\beq
\cK(s, \vec x ; t, \vec r) = 
\int {\rm d}^2\vec z \,
\cK (s, \vec x ; \tau_{\rm m}, \vec z)\, \cK_0(\tau_{\rm m}, \vec z ; t, \vec r)\,,
\eeq
with $\cK_0(\tau_{\rm m},\vec z;t,\vec r)$ 
being the vacuum kernel given by the Fourier transform
of \eqref{eq:KernelVac}, yielding
\beq
\begin{aligned}
\omega 
\frac{\mathrm{d} I}{\mathrm{d} \omega \mathrm{d}^2 \vec k}=
&\,\frac{2  \alpha_{\rm s}C_{\rm R} \,\omega}{(2 \pi)^4} 
\operatorname{Re}
\int_{\tau_{\rm m}}^{L} \mathrm{d} s \, n(s)
\int_{\tau_{\rm p}}^{\tau_{\rm m}} 
\frac{\mathrm{d} t}{(\tau_{\rm m}-t)^2} 
  \int \mathrm{d}^2 \vec x\,  
\mathrm{d}^2\vec y\, 
\mathrm{d}^2\vec z\,
i \left[\sigma(\vec{x}) - \sigma(\vec{y}) \right]
\\& \qquad \quad \times
\frac{\vec z \cdot (\vec y - \vec x)}{(\vec x-\vec y)^2} \exp\left\{i\frac{\omega}{2} \frac{\vec{z}^2}{\tau_{\rm m} - t} \right\} 
\cK(s,\vec x; \tau_{\rm m}, \vec z) \,
e^{-\frac{1}{2} \int_s^L \mathrm{d}\xi n(\xi)\sigma(\vec y)}
e^{-i\vec k \cdot \vec y}\,
\\& +
\frac{2 \alpha_{\rm s}C_{\rm R}}{(2 \pi)^3 \omega} 
\operatorname{Re}
\int_{\tau_{\rm m}}^L \mathrm{d} s \, n(s)
\int_{\tau_{\rm m}}^s \mathrm{d} t 
\int \mathrm{d}^2\vec x \,
\mathrm{d}^2\vec y \,i
\left[\sigma(\vec{x}) - \sigma(\vec y) \right]
\\&\qquad \quad \times
\frac{\vec x - \vec y}{(\vec x-\vec y)^2}  
\cdot\partial_{\vec r}
\cK \left.(s, \vec x ; t, \vec r)\right|_{\vec r=0}
e^{-\frac{1}{2}\int_s^L \mathrm{d} \xi n(\xi)\sigma(\vec{y})}
e^{-i\vec k\cdot\vec{y}}\,.
\end{aligned}
\label{eq:kt_spectrum_xspace3}
\eeq
The second term in \eqref{eq:kt_spectrum_xspace3} corresponds to the spectrum in scenarios where the highly energetic parton is produced at the same time as the starting point of the medium ($\tau_{\mathrm p}=\tau_{\rm m}$), while the first term corresponds  to the additional radiation due to the initial propagation through vacuum. 
% It is straightforward to check that this first term is zero for $\tau_{\mathrm p}=\tau_{\rm m}$.

%%%%%%%%%%%%%%%%%%%%%%%%%%%%%%%%%%%%%%%%%%%%%%%%%%%%%%%%%%%%%%%%%%%%%%%%%%%%%%%%%%%%%%%%%%%%
\subsection{HO spectrum in the brick}
\label{subsec:HO_brick}

Let us now restrict to a static scenario in which the medium is modeled as a ``brick'' with constant $\hat q$ given by $\hat q (s) = \hat{q}_0$ for $s \,\in\, [\tau_{\rm m},L]$. In this static case,  the HO kernel has an analytical expression, which is given by \cite{Salgado:2003gb,Arnold:2008iy}\footnote{
We show in the Appendix~\ref{ap:HO} the expressions of the HO kernel and spectrum for a dynamic medium where $\hat q (s)$ follows a power-law behavior. For other longitudinally evolving media profiles see for instance \cite{Arnold:2008iy,Adhya:2019qse}.}
\beq
\cK^{\mathrm{HO}}(s, \vec x ; t, \vec r) =
\frac{-i\omega\Omega}
{2\pi\sin\left[\Omega(s-t) \right]}
\exp\left\{\frac{i\omega\Omega}
{2\sin\left[\Omega(s-t) \right]}
\left(\cos\left[\Omega(s-t) \right]\vec x^2 + 
\cos\left[\Omega(s-t) \right]\vec r^2
- 2\vec x\,\cdot \vec r \right)
\right\}\,,
\label{eq:HO_kernel_brick}
 %\eqref{eq:HO_kernel_brick}
\eeq
where we have defined
\beq
\Omega \equiv \frac{ (1-i)}{2}
\sqrt{\frac{\hat{q}_0}{\omega}}\,.
\label{eq:HO_Omega_st}
%\eqref{eq:HO_Omega_st}
\eeq
Plugging \eqref{eq:sigma_HO}~and~\eqref{eq:HO_kernel_brick} into \eqref{eq:kt_spectrum_xspace3} and performing the integrals over $\vec x$ and $\vec y$ one can show that HO $\vec k$-differential spectrum can be written as 
\beq
\begin{aligned}
\omega \frac{\mathrm{d} I^{\mathrm{HO}}}
{\mathrm{d} \omega \mathrm{d}^ 2 \vec k}=\,
&\frac{ \alpha_{\rm s}C_{\rm R}}{ \pi^ 2 \omega} \operatorname{Re} 
\int_{\tau_{\rm m}}^L \mathrm{d}s 
\int_{\tau_{\rm p}}^{\tau_{\rm m}} \mathrm{d} t \,
\frac{i\hat{q}_0 }
{\left[\widetilde{A}(s,t)\right]^2} 
\frac{1}{\hat{q}_0(L-s)-2i\omega\Omega\frac{\widetilde{A}(s,t)}{\widetilde{B}(s,t)}} 
\exp\left\{\frac{-k^2}{\hat{q}_0(L-s)-2i\omega\Omega\frac{\widetilde{A}(s,t)}{\widetilde{B}(s,t)}} \right\}
 \\
&+\frac{\alpha_ s C_ R}{ \pi^2 \omega} 
\operatorname{Re} 
\int_{\tau_{\rm m}}^L \mathrm{d} s 
\int_{\tau_{\rm m}}^s \mathrm{d} t \,
\frac{i\hat{q}_0 \sec^2[\Omega(s-t)]}{\hat{q}_0(L-s)-2i\omega \Omega \cot[\Omega(s-t)]}
\exp{\left\{\frac{-k^2}
{\hat{q}_0(L-s)-2i\omega \Omega \cot[\Omega(s-t)] } \right\}} \,,
\end{aligned}
%\eqref{eq:HO_ktspec_brick}
\label{eq:HO_ktspec_brick}
\eeq
where we have defined
\beq
\widetilde{A}(s,t) \equiv \cos [\Omega (s-\tau_{\rm m})] - \Omega(\tau_{\rm m}-t)\sin [\Omega (s- \tau_{\rm m})]\,,
\label{eq:At_function}
\eeq
\beq
\widetilde{B}(s,t) \equiv \sin [\Omega (s-\tau_{\rm m})] + \Omega(\tau_{\rm m}-t)\cos [\Omega (s-\tau_{\rm m})]\,.
\label{eq:Bt_function}
\eeq

For the HO energy energy distribution one needs to integrate \eqref{eq:HO_ktspec_brick} over $\vec k$ with $0 \leq k \leq \omega$ (considering the kinematic constraint) and rescale the time 
variables as $t \rightarrow t/L$ and $s \rightarrow s/L$  to obtain 
\beq
\begin{aligned}
\omega \frac{\mathrm{d}I^{\mathrm{HO}}}
{\mathrm{d}\omega }&=
-\frac{2\alpha_{\rm s}C_{\rm R}}{\pi}
\operatorname{Re} 
\int_{\tau_{\rm m}/L}^1 \mathrm{d}s 
\int_{\tau_{\rm p}/L}^{\tau_{\rm m}/L} \mathrm{d}t \,
\frac{i\omega_{\rm c}}{\omega A^2(s,t)} 
\left[\exp \left\{
\frac{-R \omega^2}
{2\omega^2_c\left(1-s -
\frac{i\omega \Omega LA(s,t)}
{\omega_{\rm c} B(s,t)} \right)}
\right\} -1\right]\\
& \quad-\frac{2 \alpha_{s} C_{R}}{ \pi } 
\operatorname{Re} 
\int_{\tau_{\rm m}/L}^1 \mathrm{d} s
\int_{\tau_{\rm m}/L}^s \mathrm{d} t
\frac{i\omega_{\rm c}}{\omega \cos^2\left[\Omega L (s-t)\right]}  \left[\exp{\left\{\frac{-R\,\omega^2}{2\omega_{\rm c}^2\left[1-s -\frac{\cot\left[ \Omega L (s-t) \right]}{\Omega L}
\right]} \right\}} - 1\right]\,,
\end{aligned}
%\eqref{eq:HO_spec_rfin_brick}
\label{eq:HO_spec_rfin_brick}
\eeq
where the functions $A$ and $B$ are the rescaled versions of $\widetilde A$ and $\widetilde B$, respectively,
\beq
A(s,t) \equiv
 \cos\left[\Omega L \left(s-\tau_{\rm m}/L\right)\right]
- \Omega L \left(\tau_{\rm m}/L-t \right) 
 \sin\left[\Omega L \left(s-\tau_{\rm m}/L \right)\right]\,,
\label{eq:A_function}
\eeq
\beq
B(s,t) \equiv 
\sin\left[\Omega L \left(s- \tau_{\rm m}/L \right)\right]
+ \Omega L \left( \tau_{\rm m}/L-t \right) \cos\left[\Omega L \left(s-\tau_{\rm m}/L \right)\right]\,,
\label{eq:B_function}
\eeq
and we have defined
\beq
\omega_{\rm c} \equiv \frac{1}{2}\hat{q}_0L^2\,,
\quad \mathrm{and} \quad R \equiv \omega_{\rm c} L\,.
\label{eq:HO_vairables}
\eeq
Taking the $R \rightarrow \infty$ limit with fixed $\omega_{\rm c}$ in \eqref{eq:HO_spec_rfin_brick} is equivalent to performing the integration over the transverse momentum of the emitted gluon up to infinity. In that case we get 
\beq
\omega \frac{\mathrm{d}I^{\mathrm{HO}}}
{\mathrm{d} \omega}
=\frac{ 2\alpha_{\rm s}C_{\rm R}}{\pi}
\ln \left|
\cos \left [\Omega L \left(1-\frac{\tau_{\rm m}}{L} \right) \right] - \Omega L \,\frac{\tau_{\rm m} -\tau_{\rm p}}{L}
\sin \left[\Omega L \left(1-\frac{\tau_{\rm m}}{L} \right)\right]
\right| \,.
\label{eq:HO_spec_rinf_brick}
%\eqref{eq:HO_spec_rinf_brick}
\eeq
It is worth noticing that  $\Omega L$ can be written as  function of $\omega/\omega_{\rm c}$ only, namely
\beq
\Omega L = (1-i) \sqrt{\frac {\omega_{\rm c}}{2\omega}}\,,
\eeq
which implies that the energy spectrum without the kinematical constraint in \eqref{eq:HO_spec_rinf_brick} is a function of  $\omega/\omega_{\rm c}$ and $\tau_{\rm m}/L$ only (when $\tau_{\mathrm p}=\tau_{\rm m}$ or $\tau_{\rm p}=0$ as it is always the case in this paper). Notice that accounting for the kinematic constraint only adds $R$ to the list of  independent parameters.

From \eqref{eq:HO_spec_rinf_brick}, one can  see that the energy spectrum in the general case, where the hard parton is created before entering the medium (see lower panel of figure~\ref{fig:picture_cases}), can be written in terms of the spectrum for
emitter created inside the medium (see top panel of figure~\ref{fig:picture_cases}) as
\beq
\left.\omega \frac{\mathrm{d} I^{\mathrm{HO}}}
{\mathrm{d} \omega}\right|_{\tau_{\rm m}>\tau_{\rm p}=0} =
\left.\omega \frac{\mathrm{d} I^{\mathrm{HO}}}{\mathrm{d} \omega}\right|_{\tau_{\rm m}=\tau_{\rm p}=0}
-\frac{2\alpha_{\rm s}C_{\rm R}}{\pi}
\ln \left|\frac{\cos\left(\Omega L\right)}
{\cos \left[\Omega L (1-\tau_{\rm m}/L)\right] - \Omega  \,\tau_{\rm m}
\sin \left[\Omega L(1-\tau_{\rm m}/L)\right] } \right|\,,
\label{eq:HO_difference}
%\eqref{eq:HO_difference}
\eeq
where, contrary to the GLV result, the second term in the r.h.s. does not correspond to the spectrum for a highly energetic parton propagating in a medium going from $\tau_{\rm p}=0$ to $\tau_{\rm m}$. We notice now that the numerator in the second term of the r.h.s of \eqref{eq:HO_difference} can be written as
\beq
\cos \left(\Omega L\right) 
= \cos \left[ \Omega L \left(1-\tau_{\rm m}/L \right) \right]
\cos\left( \Omega L\, \tau_{\rm m}/L \right) 
- \sin \left[ \Omega L \left(1-\tau_{\rm m}/L \right) \right]
\sin \left(\Omega L\, \tau_{\rm m}/L \right)\,,
\eeq
which for $|\Omega \tau_{\rm m}|\ll 1$ yields 
\beq
\cos\left(\Omega L\right) 
\xrightarrow[]{|\Omega \tau_{\rm m}|\ll 1} 
\cos[\Omega L (1-\tau_{\rm m}/L)] - \Omega \tau_{\rm m}
\sin \left [\Omega L (1-\tau_{\rm m}/L) \right]\,.
\eeq
Hence, the second term on the r.h.s of \eqref{eq:HO_difference} goes to $0$  for $|\Omega \tau_{\rm m}|\ll 1$, setting a new characteristic energy scale $\omega_{\rm m}$ given by
\beq
\omega_{\rm m} \equiv \frac{1}{2} \hat{q}_0  \tau_{\rm m}^2 =  \omega_{\rm c}\frac{\tau_{\rm m}^2}{L^2}\,.
\eeq
This explicitly shows that for  large gluon energies ($\omega \gg \omega_{\rm m}$) the difference between the spectra for hard partons produced inside or outside the medium becomes negligible.

\begin{figure}
\includegraphics[width=\linewidth]{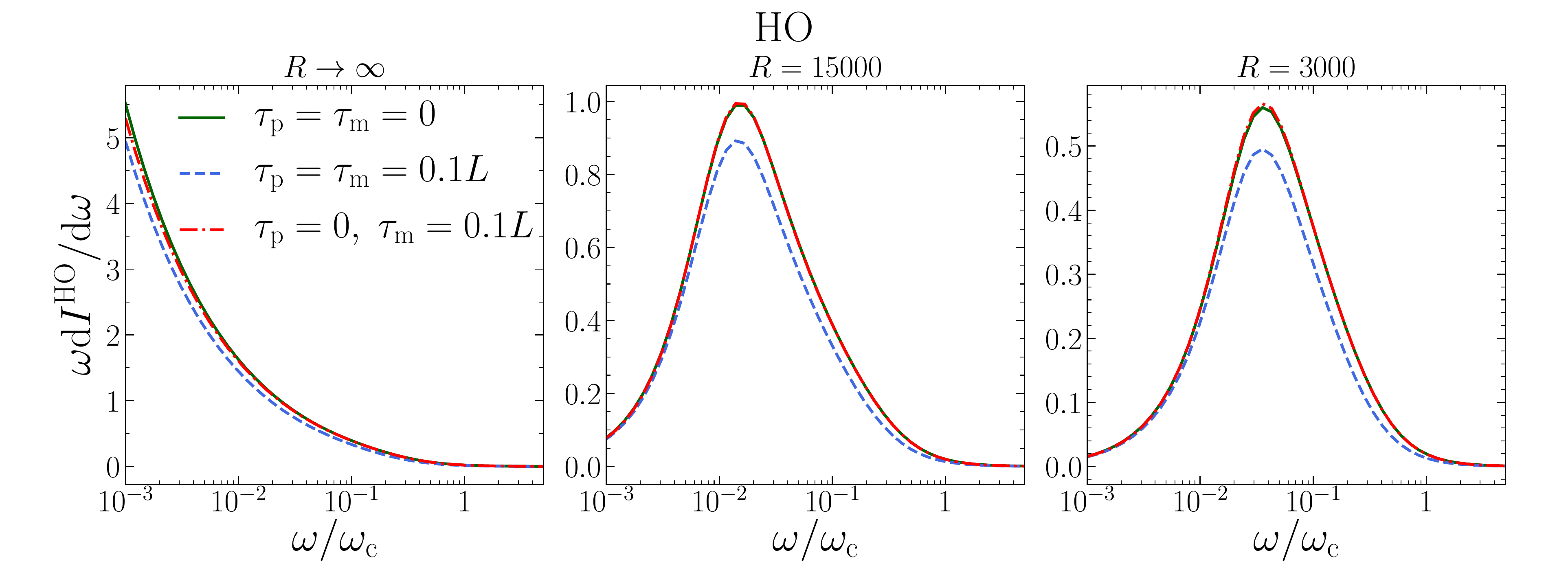}
\caption{HO medium-induced energy spectrum as a function of  $\omega/ \omega_{\rm c}$  for the three scenarios analyzed in this paper:  $\tau_{\rm p} = \tau_{\rm m} = 0$ (green solid lines);  $\tau_{\rm p} = \tau_{\rm m} =0.1 \,L $ (blue dashed lines); and  $\tau_{\rm p} = 0$ and $\tau_{\rm m}= 0.1\,L$ (red dash-dotted lines).  The different panels correspond to different sets of $R$: $R\rightarrow\infty$ (left), $R=15000$ (central), and $R=3000$ (right).}
 \label{fig:HO_t0L0_1}
\end{figure}

\begin{figure}
\centering
\includegraphics[width=\linewidth]{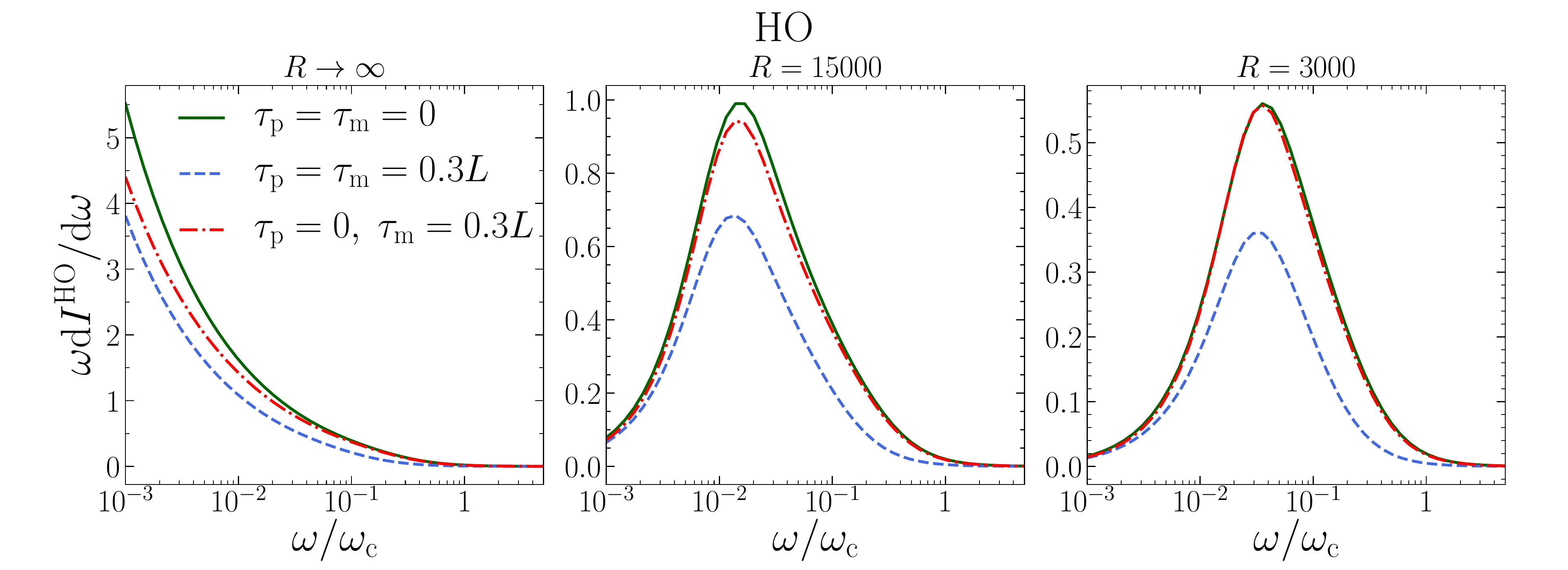}
\caption{HO medium-induced  energy spectrum as function of  $\omega/ \omega_{\rm c}$  for the three scenarios analyzed in this paper:  $\tau_{\rm p} = \tau_{\rm m} = 0$ (green solid lines); $\tau_{\rm p} = \tau_{\rm m} =0.3\,L $ (blue dashed lines); and $\tau_{\rm p} = 0$ and $\tau_{\rm m} = 0.3\,L$  (red dash-dotted lines). The different panels correspond to different sets of $R$: $R\rightarrow\infty$ (left), $R=15000$ (central) and $R=3000$ (right).}
 \label{fig:HO_t0L0_3}
\end{figure}

We present in figures~\ref{fig:HO_t0L0_1}~and~\ref{fig:HO_t0L0_3} the HO medium-induced energy spectrum off a highly energetic quark as a function of $\omega/\omega_{\rm c}=2\omega/(\hat{q}_0L^2)$ for the three initial stages scenarios analyzed in this manuscript. The medium is considered as static in both figures and we use green solid lines for $\tau_{\rm p} = \tau_{\rm m}=0$; blue dashed lines for $\tau_{\rm p} = \tau_{\rm m} > 0$; and red dash-dotted lines for $\tau_{\rm m} > \tau_{\rm p}=0$.  When different from zero, $\tau_{\rm m}$ is set to $\tau_{\rm m}/L=0.1$ in figure~\ref{fig:HO_t0L0_1} and $\tau_{\rm m}/L=0.3$ in figure~\ref{fig:HO_t0L0_3}. On the left panel of both figures we show the energy distribution without the kinematic condition ($R \rightarrow \infty$) computed using eq.~\eqref{eq:HO_spec_rinf_brick}, while the central and right panels show the energy spectrum where the kinematic constraint $k \leq \omega $ has been imposed (computed through eq.~\eqref{eq:HO_spec_rfin_brick}) for two different values of the parameter $R$: $R=15000$ (central panel) and $R=3000$ (right panel).

The results shown in figures~\ref{fig:HO_t0L0_1}~and~\ref{fig:HO_t0L0_3} follow the same pattern as those from the previous sections. The spectra for the case where the hard parton is allowed to propagate in vacuum before entering the medium ($\tau_{\rm m} > \tau_{\rm p} =0$,  red dash-dotted lines) are identical to the corresponding ones when the medium starts at the same point where the emitter is created ($\tau_{\rm p} = \tau_{\rm m}=0$, green solid lines) for large energies, while for low energies the scenario accounting for vacuum propagation becomes smaller. Nonetheless, we see that the differences between those two initial stages cases within the HO approximation are much smaller than those observed in the previous sections. This is a direct consequence of the relevant energy scale for the HO being $\omega_{\rm m}\sim (\tau_{\rm m}/L)^2$ instead of $\bar\omega_{\rm m}\sim \tau_{\rm m}/L$. The point where the red  dash-dotted and  green solid curves start to diverge is thus much smaller for the HO approach than for the GLV and full formalisms.

%%%%%%%%%%%%%%%%%%%%%%%%%%%%%%%%%%%%%%%%%%%%%%%%%%%%%%%%%%%%%%%%%%%
%%%%%%%%%%%%%%%%%%%%%%%%%%%%%%%%%%%%%%%%%%%%%%%%%%%%%%%%%%%%%%%%%%%

\section{$R_{\rm AA}$ and high-$p_T$ $v_2$}
\label{sec:v2}

Now that we know how to account for the additional parton propagation before the formation of the QGP, we address, in this section, the potential effects that this initial radiation may have on a selection of jet observables. The first natural step is to look at the one combination of jet observables which has already been shown to be sensitive to the initial stages after the collision. Although the simultaneous determination of the single-inclusive particle suppression $R_{\rm AA}$ and the high-$p_T$ azimuthal asymmetry $v_2$ has been subject of numerous studies in the last decade, see for instance \cite{Noronha-Hostler:2016eow,Betz:2012qq,Jia:2011pi,Renk:2010qx,Liao:2008dk} and references therein, it was not pointed out until very recently that their combined description is very sensitive to the starting point of the quenching  \cite{Andres:2019eus}. In fact, all phenomenological studies which successfully describe them simultaneously delay the interaction of the emitter with the medium up to the initial time of the hydrodynamic simulation \cite{ Zhao:2021vmu,Noronha-Hostler:2016eow,Betz:2016ayq,Zigic:2018ovr,Zigic:2019sth}, usually taken as\footnote{We note that this is the range employed up to now in jet quenching studies, but recent studies have shown that an earlier transition to the hydrodynamic regime at $\tau_{\rm hydro} \sim 0.2$\,fm, is also possible \cite{Nijs:2021clz}.} $\tau_{\rm hydro} = 0.6-1$\,fm.\footnote{ Alternatively, in \cite{Shi:2018lsf,Shi:2018izg} a magnetic monopole was introduced producing a large enhancement of $\hat q$ for temperatures close to the deconfinement temperature, whose effect is very similar to delaying the interaction of the probe with the medium.}

Our main goal is not to have an accurate description of the experimental data but to illustrate  the effects induced by the additional initial radiation when attempting a simultaneous description and interpretation of the $R_{\rm AA}$ and high-$p_T$ $v_2$. For this purpose, we follow the same procedure for simultaneously obtaining the two observables in each of the three scenarios considered along this paper. We use a simple model in which the hard parton first propagates through vacuum and then enters an expanding medium modeled either through a straightforward power-law expansion or a smooth-averaged hydrodynamic simulation. We then calculate the quenching experienced by the hard parton in specific trajectories (selected according to their in-medium length) which allows us to approximate the overall suppression and the corresponding azimuthal asymmetry.

\subsection{The formalism}

As a measure of the suppression for a given parton species $i$, in a fixed trajectory, we will use the quenching factor $Q_i(p_T)$ defined as the ratio of the partonic medium and vacuum spectra
\beq
Q_i(p_T)= 
\frac{{\rm d}\sigma^{{\rm AA}}(p_T)/{\rm d}p_T}
{{\rm d}\sigma^{{\rm pp}}(p_T)/{\rm d}p_T} = 
\int {\rm d}\epsilon \, P(\epsilon) 
\frac{{\rm d}\sigma^{{\rm pp}}(p_T + \epsilon)/{\rm d}p_T}{{\rm d}\sigma^{{\rm pp}}(p_T)/{\rm d}p_T}
\,,
\label{eq:quenching_factor}
%\eqref{eq:quenching_factor}
\eeq
where $P(\epsilon)$ are the quenching weights (QWs) which encode the probability that a highly energetic parton loses an amount of energy $\epsilon$ due to the radiation of an arbitrary number of independent medium-induced gluons \cite{Salgado:2003gb}. The QWs are defined as \cite{Salgado:2002cd,Salgado:2003gb}
\beq
P(\epsilon)= 
\sum_{N=0}^\infty \frac{1}{N!} \,
\prod_{i=1}^N \,
\left [\int {\rm d}\omega_i \,\frac{{\rm d}I}{ {\rm d} \omega} \right]
\,\delta \left(\epsilon -\sum_{i=1}^{N} \omega_i \right)
\exp \left[-\int_0^\infty {\rm d}\omega \frac{{\rm d}I}{{\rm d}\omega} \right]\,,
\label{eq:QW}
%\eqref{eq:QW}
\eeq
where ${\rm d}I/{\rm d}\omega$ is the medium-induced energy spectrum with the kinematic cut-off imposed for a fixed trajectory. 

If we approximate the initial parton spectrum by a power law ${\rm d}\sigma^{{\rm pp}}/{ \rm d}p_T \propto p_T^{-n}$ with $n$ constant, the quenching factor \eqref{eq:quenching_factor} can be written as
\beq
Q_i(p_T)= \int {\rm d}\epsilon \, P(\epsilon) 
\left (\frac{p_T + \epsilon}{p_T} \right)^n\,.
\label{eq:quenching_factor2}
%\eqref{eq:quenching_factor2}
\eeq
In this study, we will estimate the inclusive suppression of charged particles (and use as proxy for the $R_{\rm AA}$) as the average of the quenching factor of a quark $q$ along the shortest (in-plane direction) and longest trajectories (out-of-plane direction)
\beq
Q_{\rm AA} (p_T)=
\frac{Q_q^{\rm in}(p_T) + Q_q^{\rm out}(p_T)}{2}
\,,
\label{eq:QAA}
\eeq
where $Q_q^{\rm in}(p_T)$ and $Q_q^{\rm out}(p_T)$ denote the (partonic) quenching factors of a quark $q$ along the  direction of the event plane and along the direction perpendicular to the event plane, respectively.

The azimuthal asymmetry in the hard sector can be approximated in a simple model by \cite{Zigic:2018smz,Adhya:2021kws}
\beq
w_2 = \frac{1}{2} 
\frac{
Q_q^{\rm in}(p_T) - Q_q^{\rm out}(p_T)}{Q_q^{\rm in}(p_T) + Q_q^{\rm out}(p_T)}\,,
\label{eq:v2}
\eeq
which we employ as proxy for the high-$p_T$ $v_2$.

For the remainder of this section we will first compute the quenching factor $Q_{\rm AA}$ of a highly energetic quark using either the HO energy spectrum derived in section~\ref{sec:HO} or the full result presented in section~\ref{sec:full} within the three initial stages scenarios shown in figure~\ref{fig:picture_cases}. The medium parameters are parameterized in terms of the local temperature and the  remaining free parameter fitted to ALICE data on the charged hadron $R_{\rm AA}$ for the $20\text{-}30\%$  centrality class in $\sqrt{s_{\mathrm{NN}}}=2.76\,{\rm TeV}$ Pb-Pb collisions \cite{ALICE:2012aqc}. This procedure will be made explicit for each of the approaches in sections \ref{subsec:HO} and \ref{subsec:full}, respectively. This fitted-value is then employed to obtain the high-$p_T$ azimuthal asymmetry $w_2$ for the $20\text{-}30\%$ centrality class in $\sqrt{s_{\rm NN }}=2.76$\,TeV Pb-Pb collisions.
 
\subsection{Results within the HO approach} 
\label{subsec:HO}

Taking into account that the formalism of the QWs, and its phenomenological applications, has been mostly developed in the HO approximation, we start by obtaining the high-$p_T$ $Q_{\rm AA}$ and $w_2$ within this setup, described in section~\ref{sec:HO}. The highly energetic quark is produced at a time $\tau_{\rm p}$ and propagates in a straight-line trajectory along either the in-plane or out-of-plane direction. In order to obtain the radiated energy spectrum one needs to specify the relation between the local transport coefficient $\hat{q}(\xi)$ and the local value of the temperature, at a given time $\xi$, which we set to
\beq
\hat{q}(\xi) = K_1\,T^{3}(\xi)\,,
\label{eq:qhat}
%\eqref{eq:qhat}
\eeq
motivated by \cite{Baier:2002tc}, in the QGP phase. Note that we do not consider in this manuscript any quenching in the hadronic phase of the system.  We then model the temperature along the path as 
\beq
T(\xi) = T_0\left(\frac{\xi_0}{\xi+\xi_0}\right)^c\,,
\label{eq:temp_HO}
%\eqref{eq:temp_HO}
\eeq
to take advantage of the fact that analytical solutions exist for the HO energy distribution when $\hat{q}(\xi)$ follows a power-law profile, as shown in Appendix~\ref{ap:HO}. The parameters $\xi_0$, $T_0$ and $c$ entering \eqref{eq:temp_HO} are determined by comparing to the temperature profiles along straight line trajectories within the 2+1 viscous hydrodynamic model for $\sqrt{s_{\rm NN}} = 2.76~\rm{TeV}$ Pb-Pb collisions given in \cite{Luzum:2008cw,Luzum:2009sb}, which we have checked can be reasonably well approximated by a power-law expansion. For completeness, we summarize here the basic features of this hydrodynamic simulation and refer the reader to  \cite{Luzum:2008cw, Luzum:2009sb} for further details. This model starts at an initial proper time of $\tau_ {\rm hydro}=1\,\rm{fm}$, takes its initial condition from a factorized Kharzeev-Levin-Nardi model \cite{Drescher:2006pi} and has the ratio of shear viscosity to entropy density fixed to a constant value of $\eta/s=0.16$. The simulation uses an equation of state inspired by Lattice QCD calculations and the system is assumed to be in chemical equilibrium until it reaches a freeze-out temperature $T_{\rm f} = 140$\,MeV. We set $\xi_0=\tau_{\rm hydro}=1$\,fm and then use the value of the temperatures for the $20\text{-}30\%$ centrality class along the in-plane direction at the initial and final times to obtain $c^{\rm in}=0.72$ and $T_0^{\rm in}=610$\,MeV. Analogously, for the out-of-plane path we get $c^{\rm out}=0.67$ and $T_0^{\rm out}=589$\,MeV.

We make use of eq.~\eqref{eq:HO_spec_rfin_dyn} to compute the energy spectrum for the three initial stages scenarios along the in-plane and out-of-plane directions. Then, following \cite{Salgado:2003gb}, we obtain the QWs $P(\epsilon)$ along both paths, which we employ to compute the in-plane and out-of-plane quenching factors using eq.~\eqref{eq:quenching_factor2} with $n=4$. We obtain the $Q_{\rm AA}$ by means of eq.~\eqref{eq:QAA} for several values of our fitting parameter $K_1$ and perform a $\chi^2$-fit to the ALICE $R_{\rm AA}$ data for the $20\text{-}30\%$ centrality class in $\sqrt{s_{\rm NN}} = 2.76~\rm{TeV}$  Pb-Pb collisions \cite{ALICE:2012aqc} to extract its best value. Finally, the fitted $K_1$ is used to obtain the corresponding high-$p_T$ $w_2$ by means of eq.~\eqref{eq:v2}, as done in ref.~\cite{Andres:2019eus}. 

\begin{table}
\centering
\begin{tabular}{ll}
\hline
Case   & $K_1$ \\
\hline
A: $\tau_{\rm p}=\tau_{\rm m}=0$\,fm & $1.5$      \\
B: $\tau_{\rm p}=\tau_{\rm m}=1$\,fm   & $16$     \\
C: $\tau_{\rm p}=0~\mathrm{and}~\tau_{\rm m}= 1$\,fm & $8$    \\
\hline
\end{tabular}
\caption{Values of $K_1$ obtained by fitting
 ALICE $R_{\rm AA}$ \cite{ALICE:2012aqc} with $p_T > 6$ GeV data for the $20\text{-}30\%$ centrality class in  $\sqrt{s_{\rm NN}} = 2.76~\rm{TeV}$  Pb-Pb collisions for the three different cases analyzed in this paper  (see figure~\ref{fig:picture_cases}).}
\label{tab:K1_factor_HO}
\end{table}

This procedure is repeated for  the three scenarios considered: $\tau_{\rm p} = \tau_{\rm m} = 0$\,fm; $\tau_{\rm p} = \tau_{\rm m} =1$\,fm; and $\tau_{\rm m} =1$\,fm and $\tau_{\rm p} = 0$\,fm. Notice that the first case involves an extrapolation to include a non-zero temperature before the hydrodynamization time, while the other cases use as starting point for the medium $\tau_{\rm m}=\tau_{\rm hydro}$. Table~\ref{tab:K1_factor_HO} shows the fitted values of $K_1$ for each scenario\footnote{We remind the reader that in all numerical evaluations in this paper we use $C_{\rm R}=C_{\rm F} = 4/3$ and $\alpha_s=0.3$.}, making evident how much $\hat{q}$ changes depending on how the initial stages are modeled (see also ref.~\cite{Apolinario:2022vzg}). Clearly, the case with the longest medium ($\tau_{\rm p} = \tau_{\rm m} = 0$) needs a much smaller value of $\hat q$ to obtain the same amount of quenching as the other cases, while the difference between including or not the radiation in vacuum before hydrodynamization is a factor of $2$ in the fitted parameter. These outcomes clearly illustrate the strong dependence on the treatment of early times in jet quenching analyses for the  extraction of the jet quenching parameter from single-inclusive suppression observables. 

\begin{figure}
\centering
\includegraphics[width=\linewidth]{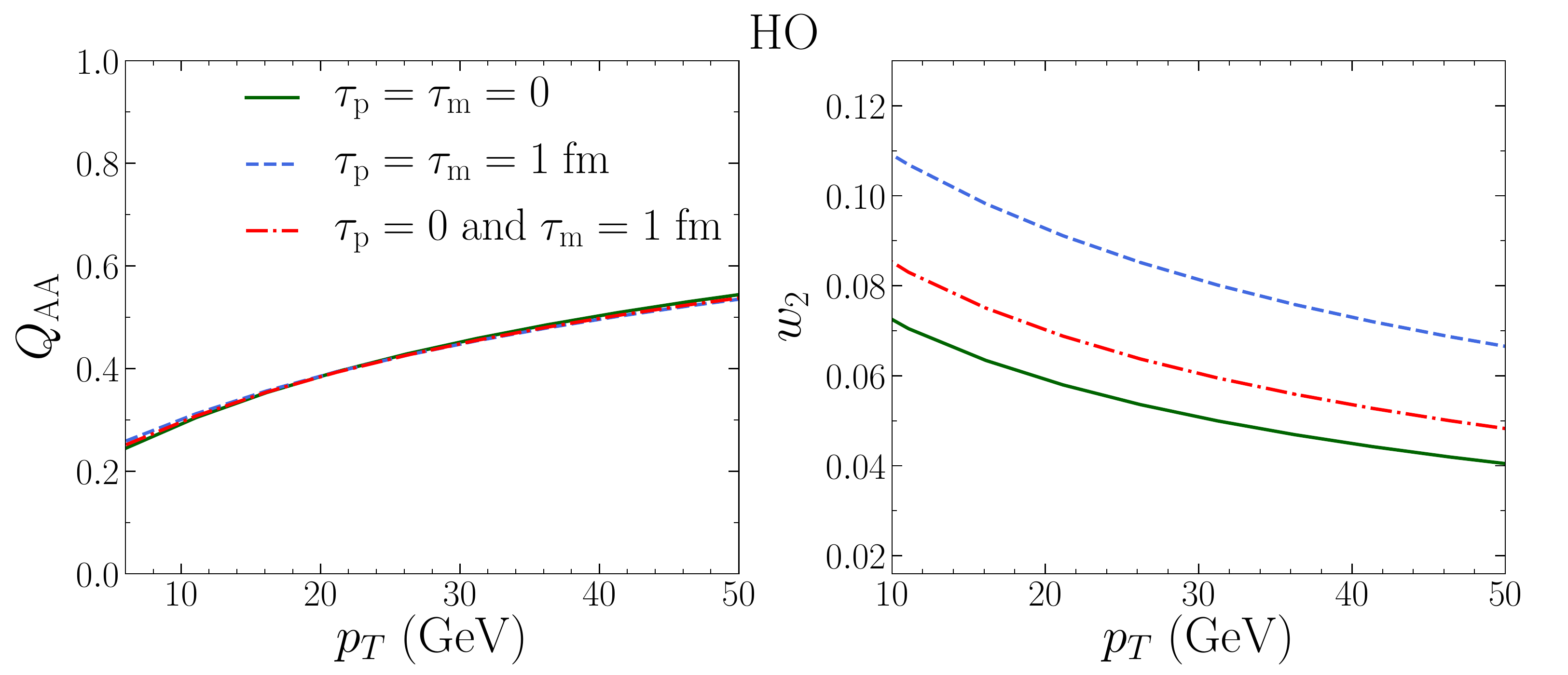}
\caption{Our proxies for the $R_{\rm AA}$ (left panel) and high-$p_T$ $v_2$ (right panel)  for the $20\text{-}30\%$ centrality class in  $\sqrt{s_{\rm NN}} = 2.76~\rm{TeV}$ Pb-Pb collisions at the LHC as function of $p_T$ obtained within the HO approximation for the three scenarios analyzed in this paper:  $\tau_{\rm p} = \tau_{\rm m} = 0$\,fm (green solid lines),  $\tau_{\rm p} = \tau_{\rm m} =1$\,fm (blue dashed lines); and $\tau_{\rm p} = 0$\,fm and $\tau_{\rm m} =1$\,fm (red dash-dotted lines). }
\label{fig:v2_HO}
\end{figure}

We show in figure~\ref{fig:v2_HO} the obtained $Q_{\rm AA}$ (left panel) and high-$p_T$ $w_2$ (right panel) as a function of $p_T$ for these three scenarios. While the fitted values of $K_1$ shown in Table~\ref{tab:K1_factor_HO} were significantly different for the three early times cases, the mild dependence of the $Q_{\rm AA}$ on the initial stage scenario employed is in agreement with previous findings \cite{Armesto:2009zi,Andres:2017awo,Andres:2016iys,Andres:2019eus}. Regarding the azimuthal asymmetry in the hard sector, we find it very sensitive to the treatment of the energy loss at early times, thus corroborating the importance of correctly accounting for in-medium radiation during the initial stages. Our results are consistent with the findings of \cite{Andres:2019eus} where it was shown that starting the hydrodynamical simulation at a later time increased the asymmetry. Nevertheless, it should be noted that the correct way of removing the interactions with the medium in the early stages is by having initial propagation in vacuum (as for the red dash-dotted lines in figure \ref{fig:v2_HO}), which corresponds to a smaller increase in the asymmetry than the case where the parton is created at the hydrodynamization time (blue dashed lines).

%%%%%%%%%%%%%%%%%%%%%%%%%%%%%%%%%%%%%%%%%%%%%%%%%%%%%%%%%%%%%%%%%%%%%%%%%%%%%%%%
%%%%%%%%%%%%%%%%%%%%%%%%%%%%%%%%%%%%%%%%%%%%%%%%%%%%%%%%%%%%%%%%%%%%%%%%%%%%%%
%% Full
\subsection{Results using the all-order spectrum} 
\label{subsec:full}

We now move to the computation of the $Q_{\rm AA}$ and $w_2$  within the fully resummed in-medium radiation formalism of section~\ref{sec:full} for the three different scenarios shown in figure~\ref{fig:picture_cases}. In order to compute the emitted energy distribution along the quark's path, we establish the relation between the medium parameters entering the fully resummed formalism ($\mu$ and $n$) and the local value of the hydrodynamic variables to

\beq
n(\xi) = C_1\, T(\xi)\,,\quad \mathrm{and} \quad\mu^2(\xi) = \frac{6\pi \alpha_s}{e} T^2(\xi)\,,
%\eqref{eq:full_param}
\label{eq:full_param}
\eeq
where we take the local temperature $T(\xi)$ directly from \cite{Luzum:2008cw, Luzum:2009sb} and keep $\alpha_s = 0.3$. For the screening mass $\mu$ we have taken the Debye mass $m_{\rm D}$ at leading order in the hard thermal loop (HTL) formalism \cite{Aurenche:2002pd} and used the relation $\mu^2 = m_{\rm D}^2/e$ allowing us to match the Yukawa and HTL cross-sections at leading logarithmic accuracy (see \cite{Barata:2020sav,Andres:2020vxs}). Since the all-order spectrum always requires a numerical evaluation, there is no advantage in parametrizing the medium temperature time evolution. The exception is the case where we take $\tau_{\rm m}=0$  (top panel of figure~\ref{fig:picture_cases}), and thus have to extrapolate the temperature before the initial time of the hydrodynamic simulation. For times prior to $\tau_{\rm hydro}$ we assume the temperature to follow the power-law profile \eqref{eq:temp_HO}. With these ingredients we can now compute the fully resummed in-medium spectrum along the in-plane and out-of-plane trajectories in the three initial stages scenarios considered. We note that the numerical evaluation of the full emission spectrum is performed following the same procedure employed in section~\ref{sec:full} and  ref.~\cite{Andres:2020vxs} for static media. Although this method is still valid for longitudinally expanding media, as the one employed here, it requires to evaluate $n(\xi)$ and $\mu^2(\xi)$ at each point of the time-evolution of the propagators, and thus the computational time needed to solve their differential equations increases substantially. We then employ these spectra to obtain the $Q_{\rm AA}$ with $n=4$ for several values of the parameter $C_1$. We  perform a $\chi^2$-fit to the ALICE $R_{\rm AA}$ data  for the $20\text{-}30\%$ centrality class  in $\sqrt{s_{\rm NN}} = 2.76~\rm{TeV}$ Pb-Pb collisions \cite{ALICE:2012aqc} to extract the best value of $C_1$ for each initial stage scenario, which is then used to obtain the corresponding high-$p_T$ $w_2$.

\begin{table}
\centering
\begin{tabular}{ll}
\hline
Case   & $C_1$ \\
\hline
A: $\tau_{\rm p}=\tau_{\rm m}=0$\,fm    & $2.2$      \\
B: $\tau_{\rm p}=\tau_{\rm m}=1$\,fm   & $4.5$     \\
C: $\tau_{\rm p}=0~\mathrm{and}~\tau_{\rm m}=1$\,fm & $2.5$    \\
\hline
\end{tabular}
\caption{Values of $C_1$ obtained by fitting ALICE $R_{\rm AA}$ \cite{ALICE:2012aqc} data with $p_T > 6$\,GeV  for the $20\text{-}30\%$ centrality class in  $\sqrt{s_{\rm NN}} = 2.76~\rm{TeV}$  Pb-Pb collisions for the three different cases analyzed in this paper  (see figure~\ref{fig:picture_cases}).}
\label{tab:C1_factor_full}
\end{table}

\begin{figure}
\centering
\includegraphics[width=\linewidth]{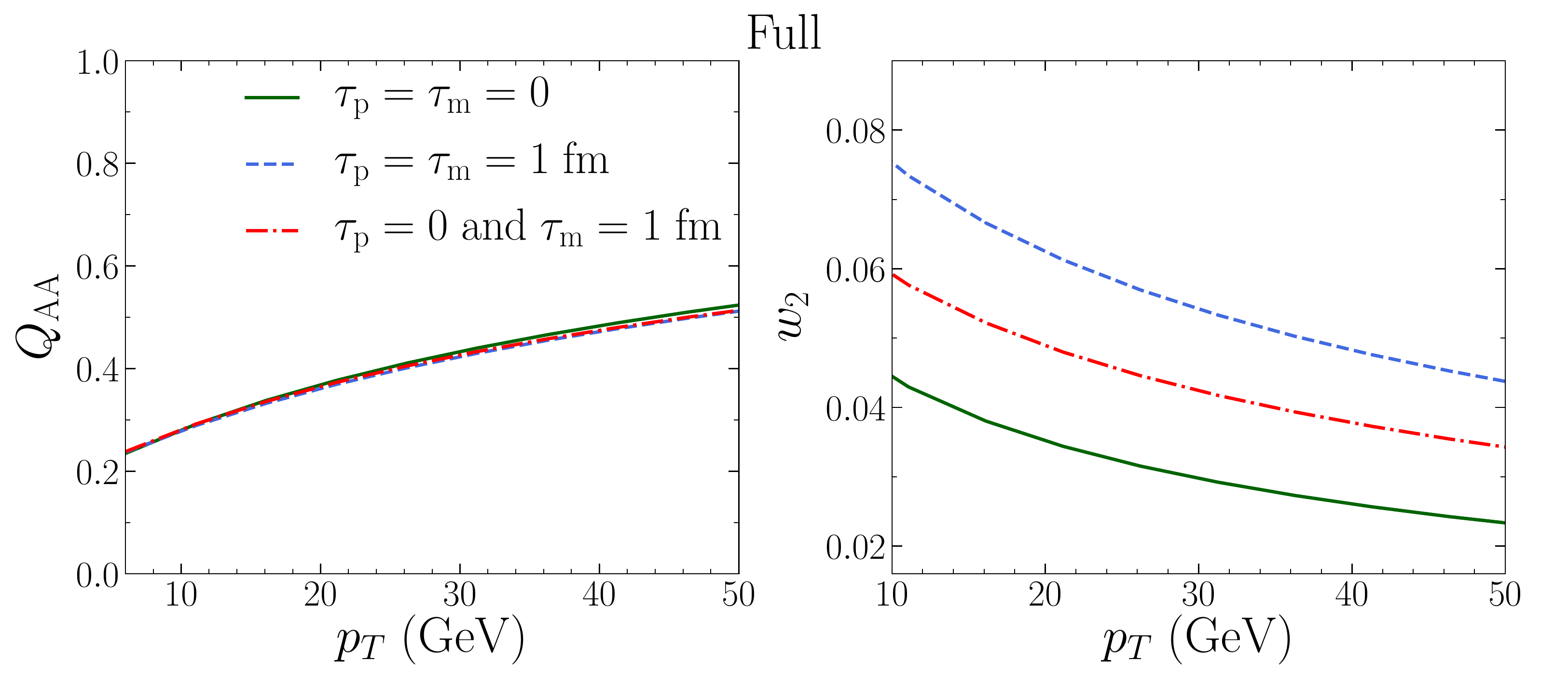}
\caption{Our proxies for the $R_{\rm AA}$ (left panel) and high-$p_T$ $v_2$ (right panel)  for the $20\text{-}30\%$ centrality class in  $\sqrt{s_{\rm NN}} = 2.76~\rm{TeV}$ Pb-Pb collisions at the LHC as function of $p_T$ obtained using the all-order BDMPS-Z spectrum for the three scenarios analyzed in this manuscript: $\tau_{\rm p} = \tau_{\rm m} = 0$\,fm (green solid lines); $\tau_{\rm p} = \tau_{\rm m} = 1$\,fm (blue dashed lines); and  $\tau_{\rm m} =1$\,fm  and $\tau_{\rm p} = 0$\,fm (red dash-dotted lines).}
\label{fig:v2_full}
\end{figure}

We present in table~\ref{tab:C1_factor_full} the fitted values of $C_1$ for the three initial stages scenarios. Even though we obtain qualitative agreement with the result of the HO case, the variation in the fitting parameter is much smaller when the full spectrum is used, suggesting that a better description of the medium interactions yields more robust results. In figure~\ref{fig:v2_full} we show the resulting $Q_{\rm AA}$ (left panel) and high-$p_T$ $w_2$ (right panel) as a function of $p_T$ for each of these scenarios. Again, the results for the full spectrum are consistent with the observations made in the HO case, but the difference between the asymmetry in the scenario with the initial propagation in vacuum (red dash-dotted line) and the scenario where the medium starts at $\tau_{\rm m} = 0$ (green solid line) is significantly larger for the full case, which is a direct consequence of the differences between the corresponding spectra being also larger, as seen in sections \ref{sec:full}~and~\ref{sec:HO}. This outcome shows for the first time the strong dependence of this set of observables on the handling of the initial stages after the collision in a formalism with full resummation of multiple scatterings with a realistic medium-probe interaction. Reaching this result within a more accurate calculation for the medium-induced radiation, which does not require the usually employed HO or GLV approximations, reveals the potential of jet quenching to constraint the dynamics of the initial stages.

\section{Conclusions and outlook}
\label{sec:conclusions}
In this manuscript we have generalized the usual BDMPS-Z formalism for medium-induced gluon radiation in which the emitter is created \emph{inside} the medium to account for its propagation and radiation before the formation of the QGP (and its interference with emissions occurring inside the medium). In order to assess the impact of this initial radiation, we assume that before the formation time of the medium $\tau_{\rm m}$ the highly energetic parton propagates as in vacuum. Within this set-up, we have derived the expression of the in-medium emission spectrum with full resummation of multiple scatterings. By comparing the numerical evaluation of this expression  with the spectra for scenarios in which the hard parton is produced inside a medium with different starting points, we have isolated the contribution from the initial stage radiation. We find that while letting the initial parton propagate in vacuum instead of in the medium in the initial part of its trajectory has almost no impact on the emission of gluons with large formation time (large $\omega$), it results in a substantial reduction of the spectrum for gluons with short formation times (see figure~\ref{fig:full_t0L0_3}). For the cases where analytical approximations are available (first opacity and harmonic oscillator approaches), we explicitly calculate the energy scale which determines the point where the spectra corresponding to initial propagation in vacuum and initial propagation in medium start to diverge.

As an illustration of the relevance of these new terms, we have employed this generalized BDMPS-Z framework to determine the contribution of this initial radiation to the simultaneous calculation of the single-inclusive hadron suppression $R_{\rm AA}$ and high-$p_T$ azimuthal asymmetry $v_2$, using as proxies the suppression factor $Q_{\rm AA}$ and the asymmetry $w_2$ defined in section~\ref{sec:v2}. We find, both for the harmonic and fully resummed approaches, that when replacing the initial slab of the in-medium propagation (prior to hydrodynamization) by vacuum and fitting the free parameter to qualitatively describe the $R_{\rm AA}$ ALICE data \cite{ALICE:2012aqc}, the azimuthal asymmetry increases substantially. Nevertheless, this increase in the asymmetry is smaller than what we would get by just computing the radiation from a parton created at the hydrodynamization time, without accounting for the initial vacuum propagation. This outcome highlights the importance of the treatment of the in-medium radiation in the initial stages for the correct description and interpretation of both observables.

Knowing that the effect of including the initial radiation before the formation of the medium goes in the correct direction to have a proper combined description of both the $R_{\rm AA}$ and high-$p_T$ $v_2$, we can possibly extend our analysis in two specific directions. On one hand, one can replace the initial propagation in vacuum by a detailed model of the interactions in the pre-hydrodynamics stage of the evolution, which presumably would have a large effect in the initial radiation according to recent results reported in  \cite{Ipp:2020mjc, Ipp:2020nfu,Carrington:2021dvw,Carrington:2022bnv}. On the other hand, one can perform a more thorough phenomenological analysis, as the one in \cite{Andres:2019eus}, now using the fully resummed spectrum and including the radiation from the initial stages to have a direct comparison to data.  Both of these developments will be addressed in separate publications.

Finally, it is worth emphasizing that understanding jet quenching in the initial stages of the system evolution in A-A collisions is crucial to understand the apparent lack of jet quenching in \emph{small systems} \cite{Citron:2018lsq} and will be necessary for the proper interpretation of jet quenching physics for intermediate A-A collisions, such as the upcoming O-O run at the LHC \cite{Brewer:2021kiv} and proposed O-O and Ar-Ar runs at RHIC by sPHENIX \cite{sphenix}.

\acknowledgments
This work is supported by European Research Council project ERC-2018-ADG-835105 YoctoLHC; by Maria de Maetzu excellence program under project CEX2020-001035-M; by Spanish Research State Agency under project PID2020-119632GB-I00 and by Xunta de Galicia (Centro singular de investigación de Galicia accreditation 2019-2022); by OE - Portugal, Fundação para a Ciência e Tecnologia (FCT) under projects EXPL/FIS-PAR/0905/2021 and CERN/FIS- PAR/0032/2021; by European Union ERDF. C.A. has received funding from the European Union’s Horizon 2020 research and innovation program under the Marie Sklodowska-Curie grant agreement No 893021 (JQ4LHC). LA was supported by OE - Portugal, FCT under contract 2021.03209.CEECIND. M.G.M. was supported by Ministerio de Universidades of Spain through the National Program FPU (grant number FPU18/01966).

\appendix
\section{HO for a power-law expanding medium}
\label{ap:HO}

In this section, we consider the harmonic oscillator approach for a longitudinal evolving medium in which the transport coefficient can be parametrized by the following power-law profile
\beq
\hat{q}(s) = 
\hat{q}_0\left(\frac{t_0}{s + t_0} \right)^\alpha\,,
\label{eq:qhat_powerlaw}
%\leqref{eq:qhat_powerlaw}
\eeq
for $s \,\in\, [\tau_{\mathrm m},L].$\footnote{Note that in practice we take $t_0=\tau_{\rm m}$.} We note that for $\alpha=0$ one should recover the static scenario discussed in section~\ref{subsec:HO_brick}. The HO emission kernel for this power-law scenario (with $\alpha \neq 2$)\footnote{For the HO emission kernel and spectrum for $\alpha =2$ see the appendix~C of \cite{Salgado:2003gb}.} can be written in coordinate space as \cite{Salgado:2003gb, Arnold:2008iy}
\beq
\cK^{\mathrm{HO}}(s, \vec x ; t, \vec{r}) = 
\frac{-i\omega}{2\pi D(s,t)}
\exp\left\{\frac{i\omega}{2D(s,t)}
\left(C_1(s,t)\,\vec {x}^2 +
C_2(s,t)\,\vec r^2 - 
2\,\vec x\cdot \vec r \right)\right\}\,,
\label{eq:HO_Kernel}
%\eqref{eq:HO_Kernel}
\eeq
where we have defined $D$,  $C_1$, and $C_2$  as
\beq
D(s,t) \equiv \frac{2\nu}{[2\nu\Omega_\alpha(t_0)]^{2\nu}}(z_sz_t)^\nu \left[I_\nu(z_s) K_\nu(z_t)
 - K_\nu(z_s) I_\nu(z_t) \right]\,,
\eeq
\beq
 C_1(s,t) \equiv z_s\left(\frac{z_t}{z_s}\right)^\nu
\left[I_{\nu-1}(z_s)K_\nu(z_t) + K_{\nu-1}(z_s)I_\nu(z_t) \right]\,,
\eeq
and
\beq
C_2(s,t) \equiv z_t\left(\frac{z_s}{z_t}\right)^\nu \left[K_\nu(z_s)I_{\nu-1}(z_t) 
+ I_\nu(z_s)K_{\nu-1}(z_t) \right]\,,
\eeq
where $I_\nu(z)$ and $K_\nu(z)$ are, respectively, the modified Bessel functions of first and second kind of order $\nu \equiv 1/(2-\alpha)$ with $z$ is given by
\beq
 z(\xi) = -2i\left|\nu\right|\Omega_\alpha(t_0)(\xi+t_0)^{\frac{1}{2\nu}}\,,
\eeq
in terms of 
\beq
\Omega_\alpha (\xi) = \frac{(1-i)}{2} \sqrt{\frac{\hat{q}_0}{\omega}\xi^\alpha}\,.
\label{eq:Omega_dyn} 
\eeq
and we have defined  $z_s\equiv z(s)$ and $z_t\equiv z(t)$. One can easily check that by setting $\alpha=0$ in \eqref{eq:HO_Kernel} one recovers  HO the static kernel given in \eqref{eq:HO_kernel_brick}. 

Plugging the HO kernel \eqref{eq:HO_Kernel} into the emission spectrum in coordinate space \eqref{eq:kt_spectrum_xspace3}, performing the integral over $\vec x$ and the Fourier transform over $\vec y$,  we get
\beq
\begin{aligned}
\omega \frac{\mathrm{d} I}
{\mathrm{d} \omega \mathrm{d}^{2} \vec k}=&
\frac{ \alpha_ s C_{\rm R}}{ \pi^ 2 \omega} 
\operatorname{Re} 
\int_{\tau_{\mathrm m}}^L \mathrm{d}s \,\hat{q}(s) 
\int_ {\tau_{\mathrm p}}^{\tau_{\mathrm m}} \mathrm{d} t\, \frac{i}
{A^2_\alpha(s,t)} 
\\ & \qquad \quad \times
\frac{1}{
\int_s^L {\rm d}\xi \, \hat{q}(\xi)
-2i\omega\Omega \frac{A_\alpha(s,t)}{B_\alpha(s,t)}}
\exp\left\{\frac{-\vec k^2}
{\int_s^L d\xi \, \hat{q}(\xi)
-2i\omega\Omega \frac{A_\alpha(s,t)}{B_\alpha(s,t)}} \right\}
\\&+
\frac{ \alpha_{\rm s}C_{\rm R}}{ \pi^2 \omega} 
\operatorname{Re} 
\int_{\tau_{\mathrm m}}^L \mathrm{d}s \,\hat{q}(s)
\int_{\tau_{\mathrm m}}^s \mathrm{d}t\,
\frac{i}
{C_1^2(s,t) F_\alpha(s,t)
} 
\exp{\left\{ \frac{-\vec k^2}
{F_\alpha(s,t) } 
\right\}} \,,
\end{aligned}
\label{eq:HO_kspec_dyn}
%\eqref{eq:HO_kspec_dyn}
\eeq
where we have defined
\beq
A_\alpha(s,t) \equiv C_1(s,\tau_{\mathrm m}) 
- \frac{\tau_{\mathrm m}-t}{D(s,\tau_{\mathrm m})} \left[1 - C_1(s,\tau_{\mathrm m})C_2(s,\tau_{\mathrm m})\right]\,,
\eeq
\beq
B_\alpha(s,t) \equiv \Omega\left[D(s,\tau_{\mathrm m}) + (\tau_{\mathrm m}-t) C_2(s,\tau_{\mathrm m}) \right]\,,
\eeq
and
\beq
F_\alpha(s,t) \equiv \int_s^L d\xi\,\hat{q}(\xi) - \frac{  2i\omega C_1(s,t)}{D(s,t)}\,.
\eeq
The second term in \eqref{eq:HO_kspec_dyn} is the $\vec k$-differential spectrum in scenarios where the hard parton is produced at the same time as the starting point of the medium ($\tau_{\rm p}=\tau_{\rm m}$), while the first term corresponds to the additional radiation due to the initial propagation through vacuum. It is clear that this first term is zero for $\tau_{\rm p}=\tau_{\rm m}$. We note that by setting $\alpha=0$ in \eqref{eq:HO_kspec_dyn} one recovers the static $\vec k$-differential spectrum \eqref{eq:HO_ktspec_brick}.

Now, performing the integration over $\vec k$ in \eqref{eq:HO_kspec_dyn} with the kinematic condition $0\leq k\leq \omega$ imposed, we obtain 
\beq
\begin{aligned}
\omega \frac{\mathrm{d}I^{\mathrm{HO}}}
{\mathrm{d}\omega }
&=\frac{\alpha_{\rm s}C_{\rm R}}{\pi \omega}
\operatorname{Re} 
\int_{\tau_{\mathrm m}}^L \mathrm{d}s \,\hat{q}(s)
\int_{\tau_{\mathrm p}}^{\tau_{\mathrm m}} \mathrm{d}t\,
\frac{i}{A_\alpha^2(s,t)} 
\left[1 - \exp\left\{\frac{- \omega^2}{\int_s^L {\rm d}\xi \, \hat{q}(\xi)
-2i\omega\Omega \frac{A_\alpha(s,t)}{B_\alpha(s,t)}
}\right\} \right]
\\& +
\frac{ \alpha_{\rm s}C_{\rm R}}{ \pi \omega}
\operatorname{Re} 
\int_{\tau_{\mathrm m}}^L \mathrm{d} s \,\hat{q}(s)
\int_{\tau_{\mathrm m}}^s \mathrm{d} t \,
\frac{i}{C_1^2(s,t)} 
\left[1- \exp{\left\{\frac{-\omega^2}
{\int_s^L \mathrm{d}\xi\hat{q}(\xi) -
\frac{2i\omega C_1(s,t)}{D(s,t)}}
\right\}} \right] \,.
\end{aligned}
\label{eq:HO_spec_rfin_dyn}
\eeq
which for $\alpha=0$ and rescaling the time variables as $t\rightarrow t/L$ yields, as expected, \eqref{eq:HO_spec_rfin_brick}.

Finally, when we allow the integration over $k$ to run up to infinite, we get
\beq
\begin{aligned}
\omega \frac{\mathrm{d}I^{\mathrm{HO}}}
{\mathrm{d}\omega }
=\frac{\alpha_{\rm s}C_{\rm R}}{\pi \omega}
\operatorname{Re} 
\int_{\tau_{\mathrm m}}^L \mathrm{d}s \, \hat{q}(s) 
\left [ \int_{\tau_{\mathrm p}}^{\tau_{\mathrm m}} \mathrm{d}t \,
\frac{i}{A_\alpha^2(s,t)} +
\int_{\tau_{\mathrm m}}^s \mathrm{d} t \,
\frac{i}{C_1^2(s,t)}\right]  \,.
\end{aligned}
\label{eq:HO_spec_rinf_dyn}
%\eqref{eq:HO_spec_rinf_dyn}
\eeq
where the first term in brackets  corresponds to the additional radiation due to the initial propagation in vacuum. This term clearly vanishes when the hard parton is produced at the starting point of the medium ($\tau_{\mathrm p}=\tau_{\mathrm m}$).

\bibliographystyle{JHEP}

\bibliography{references}

\end{document}